\documentclass[aps,prb,superscriptaddress,reprint]{revtex4-1}

\usepackage{graphicx}
\usepackage{dcolumn}
\usepackage{bm}
\usepackage{amssymb}
\usepackage{amsmath}
\usepackage{subfigure}
\usepackage{color}
\usepackage[normalem]{ulem}
\usepackage[colorlinks=true, urlcolor=blue, pdfborder={0 0 0}]{hyperref}
\hypersetup{
linkcolor=blue,
citecolor=blue
}

\begin{document}

\widetext
\title{Impact of disorder on the optoelectronic properties of GaN$_y$As$_{1-x-y}$Bi$_x$ alloys and heterostructures}


\author{Muhammad Usman}
\email{musman@unimelb.edu.au} 
\affiliation{School of Physics, University of Melbourne, Parkville, Melbourne, Victoria 3010, Australia}

\author{Christopher A.~Broderick}
\email{c.broderick@umail.ucc.ie} 
\affiliation{Tyndall National Institute, Lee Maltings, Dyke Parade, Cork T12 R5CP, Ireland}
\affiliation{Department of Electrical and Electronic Engineering, University of Bristol, Bristol BS8 1UB, U.K.}

\author{Eoin P.~O'Reilly}
\affiliation{Tyndall National Institute, Lee Maltings, Dyke Parade, Cork T12 R5CP, Ireland}
\affiliation{Department of Physics, University College Cork, Cork T12 YN60, Ireland}



\begin{abstract}

We perform a systematic theoretical analysis of the nature and importance of alloy disorder effects on the electronic and optical properties of GaN$_{y}$As$_{1-x-y}$Bi$_{x}$ alloys and quantum wells (QWs), using large-scale atomistic supercell electronic structure calculations based on the tight-binding method. Using ordered alloy supercell calculations we also derive and parametrise an extended basis 14-band \textbf{k}$\cdot$\textbf{p} Hamiltonian for GaN$_{y}$As$_{1-x-y}$Bi$_{x}$. Comparison of the results of these models highlights the role played by short-range alloy disorder -- associated with substitutional nitrogen (N) and bismuth (Bi) incorporation -- in determining the details of the electronic and optical properties. Systematic analysis of large alloy supercells reveals that the respective impact of N and Bi on the band structure remain largely independent, a robust conclusion we find to be valid even in the presence of significant alloy disorder where N and Bi atoms share common Ga nearest neighbours. Our calculations reveal that N- (Bi-) related alloy disorder strongly influences the conduction (valence) band edge states, leading in QWs to strong carrier localisation, as well as inhomogeneous broadening and modification of the conventional selection rules for optical transitions. Our analysis provides detailed insight into key properties and trends in this unusual material system, and enables quantitative evaluation of the potential of GaN$_{y}$As$_{1-x-y}$Bi$_{x}$ alloys for applications in photonic and photovoltaic devices.

\end{abstract}


\maketitle


\section{Introduction}
\label{sec:introduction}

Over the past several decades significant research effort has been dedicated to the development of III-V semiconductor alloys and quantum-confined heterostructures such as quantum wells \cite{Arai_IEEEJSTQE_2013,Nakao_IEEEJSTQE_2015,Bogusevschi_IEEEJQE_2016, Berger_AIPA_2015,Lammers_APL_2016,Fuchs_EL_2016} and quantum dots, \cite{Michler_book_2017,Wu_book_2014,Bimberg_IEEEJSTQE_1997,Huffaker_APL_1998} as platforms for the development of a range of photonic, photovoltaic, and spintronic devices. Despite the widespread of use InP-based quantum well (QW) based devices in optical communications, an important factor limiting overall device performance is the prevalence of temperature-dependent loss mechanisms, including carrier leakage, as well as non-radiative Auger recombination and inter-valence band absorption (IVBA) processes involving transitions between the highest energy valence bands (VBs) and the spin-split-off (SO) band. \cite{Phillips_IEEEJSTQE_1999,Sweeney_PSSB_1999,Houle_IEEEJQE_2005} As these loss mechanisms are governed by the fundamental parameters of material band structure, in particular the band gap ($E_{g}$) and VB spin-orbit splitting energy ($\Delta_{\scalebox{0.6}{\textrm{SO}}}$), any attempt to mitigate them must focus on developing materials and heterostructures whose electronic properties can be flexibly engineered. \cite{Adams_IEEEJSTQE_2011,Broderick_SST_2012,Marko_SSMS_2014}

Recently, dilute bismide alloys have emerged as a promising new material system whose band structure can be engineered in order to suppress IVBA and Auger recombination. \cite{Li_book_2013,Broderick_SST_2012} Dilute bismides are formed when a dilute fraction of Bi atoms replace As in (In)GaAs, forming the (In)GaAs$_{1-x}$Bi$_{x}$ alloy. The incorporation of Bi leads to an extremely strong reduction and composition-dependent bowing of $E_{g}$, of $\approx 90$ meV between $x = 0$ and 1\% in GaAs$_{1-x}$Bi$_{x}$. Contrary to N incorporation in GaAs, which strongly perturbs the conduction band (CB) structure leading to a reduction of $E_{g}$ of $\approx 150$ meV between $y = 0$ and 1\% in GaN$_{y}$As$_{1-y}$, \cite{Shan_PRL_1999,Kent_PRB_2001,Reilly_SST_2009} Bi -- being significantly larger and more electropositive than As or N -- primarily impacts the VB. \cite{Janotti_PRB_2002,Zhang_PRB_2005,Deng_PRB_2010,Usman_PRB_2011} As a result, the strong reduction of $E_{g}$ in GaAs$_{1-x}$Bi$_{x}$ is accompanied by a similarly large increase and bowing of $\Delta_{\scalebox{0.6}{\textrm{SO}}}$. \cite{Fluegel_PRL_2006,Usman_PRB_2011,Batool_JAP_2012} This brings about the potential to engineer the band structure to achieve $\Delta_{\scalebox{0.6}{\textrm{SO}}} > E_{g}$, and to hence facilitate suppression of the dominant Auger recombination and IVBA mechanisms. \cite{Broderick_SST_2012,Sweeney_JAP_2013,Broderick_IEEEJSTQE_2015} Given the strong and complementary nature of the impact of N and Bi incorporation on the GaAs band structure, co-alloying to form the quaternary dilute nitride-bismide alloy GaN$_{y}$As$_{1-x-y}$Bi$_{x}$ provides significant opportunities for band structure engineering. \cite{Janotti_PRB_2002,Broderick_SST_2012,Sweeney_JAP_2013} N (Bi) incorporation primarily impacts the CB (VB) structure and introduces tensile (compressive) strain with respect to a GaAs substrate, suggesting that the band gap, VB structure, band offsets, and strain can all be readily engineered. Initial experimental studies \cite{Yoshimoto_JJAP_2004,Huang_JJAP_2004,Tixier_APL_2005,Bushell_JCG_2014} have revealed the expected giant reduction and bowing of $E_{g}$, confirming that GaN$_{y}$As$_{1-x-y}$Bi$_{x}$ alloys offer an interesting platform from the perspective of photonic and photovoltaic device development.

To fully exploit the novel characteristics of GaN$_{y}$As$_{1-x-y}$Bi$_{x}$, a comprehensive theoretical understanding of the properties of this emerging semiconductor alloy must be developed. Given the highly-mismatched nature of N- and Bi-containing semicondutor alloys -- in which substitutional N and Bi atoms act as isovalent impurities, generating localised states which hybridise with the extended (Bloch) states of the host matrix semiconductor leading to a breakdown in Bloch character -- conventional theoretical approaches to model alloy band structure, such as the virtual crystal (VC) approximation, break down (as is the case, to a lesser extent, even in more conventional semicondutor alloys \cite{Zhang_PRL_2008}). These unusual material properties then mandate that a direct, atomistic approach, free of limiting approximations, is necessary to understand the strongly perturbed electronic structure. \cite{Wei_PRL_1996,Kent_PRB_2001,Reilly_SST_2009,Zhang_PRB_2005,Usman_PRB_2011} In particular, based on previous investigations \cite{Broderick_nitride_chapter_2017,Broderick_bismide_chapter_2017} of GaN$_{y}$As$_{1-y}$ and GaAs$_{1-x}$Bi$_{x}$ alloys, it is expected that short-range alloy disorder -- associated with the formation of pairs and larger clusters of N and/or Bi atoms sharing common Ga nearest neighbours -- will have a pronounced effect on GaN$_{y}$As$_{1-x-y}$Bi$_{x}$ alloy properties. Previous studies have been primarily based on continuum approaches, which typically ignore disorder effects and therefore offer limited scope to understand the full details of the complicated alloy electronic structure, or to analyse the results of experimental measurements \cite{Nacer_OQE_2008,Sweeney_JAP_2013,Broderick_SST_2013}. Here, we present a multi-scale framework based on atomistic tight-binding (TB) and continuum \textbf{k}$\cdot$\textbf{p} models to describe the electronic and optical properties of GaN$_{y}$As$_{1-x-y}$Bi$_{x}$ alloys and heterostructures. Through systematic large-scale atomistic TB calculations, we develop a detailed understanding of GaN$_{y}$As$_{1-x-y}$Bi$_{x}$ alloys and QW structures, in particular highlighting the crucial significance of alloy disorder effects and the associated implications for the interpretation of future experimental measurements, and for development of heterostructures for device applications.

We begin our analysis by undertaking large-scale electronic structure calculations on ordered alloy supercells, and quantify the impact of co-alloying N and Bi on the GaAs electronic structure in the ultra-dilute (dilute doping) limit. Our results show that N and Bi perturb the electronic properties effectively independently of one another. For large, disordered alloy supercells we compute the evolution of the electronic structure with alloy composition, revealing general trends and demonstrating, somewhat surprisingly, that the respective impact of N and Bi incorporation on the electronic properties remains independent, even in the presence of significant short-range alloy disorder. Our analysis demonstrates that an extended basis set 14-band \textbf{k}$\cdot$\textbf{p} Hamiltonian -- which explicitly treats the localised impurity states associated with substitutional N and Bi atoms, and is parametrised directly via atomistic supercell calculations -- describes the main features of the band structure evolution with reasonable accuracy compared both to full atomistic calculations and experimental measurements. We perform atomistic and continuum calculations of the electronic and optical properties of GaN$_{y}$As$_{1-x-y}$Bi$_{x}$/GaAs QWs, the results of which highlight the role played by alloy disorder in determining the properties of technologically relevant heterostructures. On the basis of our analysis we then evaluate the potential to develop devices for practical applications, and (i) suggest, contrary to recent studies, \cite{Habchi_MSSP_2014,Song_JAP_2016,Fan_JAP_2016} that GaN$_{y}$As$_{1-x-y}$Bi$_{x}$ heterostructures are not suitable for applications at 1.55 $\mu$m, and (ii) confirm that the most promising potential application of GaN$_{y}$As$_{1-x-y}$Bi$_{x}$ alloys is as an $\approx 1$ eV band gap material, grown lattice-matched to GaAs (or Ge) for applications in multi-junction solar cells. \cite{Sweeney_IEEEPVSC_2013} Overall, our results elucidate the unusual properties of GaN$_{y}$As$_{1-x-y}$Bi$_{x}$, highlight the importance of short-range alloy disorder in determining the details of the material properties, and provide guidelines for the development of optimised photonic and photovoltaic devices based on this emerging semiconductor alloy.

The remainder of this paper is organised as follows. In Sec.~\ref{sec:theory} we describe our atomistic TB and continuum \textbf{k}$\cdot$\textbf{p} models of the GaN$_{y}$As$_{1-x-y}$Bi$_{x}$ electronic structure. Next, in Sec.~\ref{sec:results} we present our results, beginning in Sec.\ref{sec:results_dilute} with an analysis of the impact of co-alloying N and Bi on the electronic properties in the dilute doping limit, before turning in Sec.~\ref{sec:results_disordered} to analyse the evolution of the electronic structure in disordered alloys, and then the respective electronic and optical properties of GaN$_{y}$As$_{1-x-y}$Bi$_{x}$/GaAs QWs in Secs.~\ref{sec:results_qw_electronic} and~\ref{sec:results_qw_optical}. In Sec. 4 we describe the implications of the calculated trends in the electronic and optical properties for practical applications. Finally, we summarise and conclude in Sec.~\ref{sec:conclusions}.


\section{Theoretical models}
\label{sec:theory}

The unusual electronic properties of dilute nitride and bismide alloys derive from the fact that, when incorporated in dilute concentrations, N and Bi act as isovalent impurities which strongly perturb the band structure of the host matrix semiconductor. Due to the prominence of N- and Bi-related impurity effects, conventional approaches to analyse alloy band structures -- e.g.~the VC approximation -- break down, meaning that direct atomistic calculations are generally required to provide quantitative insight. \cite{Kent_SST_2002,Reilly_JPCM_2004,Reilly_SST_2009} Furthermore, since the effects of Bi and N incorporation are prominent at dilute compositions, quantitative understanding of the properties of real materials must be built on analysis of systems containing upwards of thousands of atoms, in order to mitigate finite size effects and so that there is sufficient scope to analyse important alloy disorder effects. \cite{Usman_PRB_2011, Usman_PRB_2013}

Here, we provide an overview of the atomistic TB and continuum \textbf{k}$\cdot$\textbf{p} models we have developed to study the GaN$_{y}$As$_{1-x-y}$Bi$_{x}$ electronic structure. Full details of these models, including the parameters used in our calculations, can be found in Sec.~S1 of the Supplementary Material.


\subsection{Atomistic: $sp^{3}s^{*}$ tight-binding model}
\label{sec:theory_tb}

Since the TB method employs a basis of localised atomic orbitals, it is ideally suited to probe the electronic structure of localised impurities. \cite{Reilly_JPCS_2010} This, combined with its low computational cost compared to first principles methods, means that appropriately parametrised TB models provide a physically transparent and highly effective means by which to systematically analyse the properties of large, disordered systems and realistically-sized heterostructures. We have previously demonstrated that the TB method provides a detailed understanding of the electronic and optical properties of GaN$_{y}$As$_{1-y}$ and GaAs$_{1-x}$Bi$_{x}$ alloys, and that calculations based on this approach are in quantitative agreement with a wide range of experimental data. \cite{Lindsay_PRL_2004,Reilly_SST_2009,Usman_PRB_2011,Usman_PRB_2013,Broderick_PRB_2014} Here, we extend this approach to GaN$_{y}$As$_{1-x-y}$Bi$_{x}$ alloys.

Our nearest-neighbour $sp^{3}s^{*}$ TB model for GaN$_{y}$As$_{1-x-y}$Bi$_{x}$ is closely based upon that developed in Ref.~\onlinecite{Usman_PRB_2011} for dilute bismide alloys, which we have previously employed to provide quantitative understanding of the electronic, \cite{Usman_PRB_2011} optical \cite{Usman_PRB_2013} and spin \cite{Broderick_PRB_2014} properties of GaAs$_{1-x}$Bi$_{x}$. In this model, which explicitly includes the effects of spin-orbit coupling, the orbital energies at a given atomic site are computed depending explicitly on the local neighbour environment, and the inter-atomic interactions are taken to vary with the relaxed nearest-neighbour (i) bond lengths via a generalisation of Harrison's scaling rule, \cite{Ren_PRB_1981,Harrison_PRB_1982} and (ii) bond angles via the two-centre expressions of Slater and Koster. \cite{Slater_PR_1954} To treat GaN$_{y}$As$_{1-x-y}$Bi$_{x}$ we have made one significant modification to this model, by including an on-site renormalisation that corrects the orbital energies at a given atomic site depending on the local displacement of the atomic positions due to lattice relaxation. This simple modification -- which depends only on the differences in atomic orbital energies and bond lengths between the constituent GaN, GaAs and GaBi compounds -- is motivated by our previous analysis of GaN$_{y}$As$_{1-y}$ alloys, where we found that it suitably describes the charge transfer associated with the non-local character of the change in the supercell Hamiltonian due to substitutional N incorporation. \cite{Lindsay_thesis_2002} As in Ref.~\onlinecite{Usman_PRB_2011}, the relaxed atomic positions in the alloy supercells are computed using a valence force field model based on the Keating potential. \cite{Keating_PR_1966,Lazarenkova_APL_2004} We note that this TB-based approach reproduces the detailed features of the electronic structure of N- and Bi-containing alloys compared to first principles calculations based on density functional theory. A detailed description of our theoretical framework can be found in our recently published review of the theory and simulation of dilute bismide alloys, Ref.~\onlinecite{Broderick_bismide_chapter_2017}, where the validity of this approach is evaluated in the context of first principles calculations and experimental measurements.

To study the properties of bulk GaN$_{y}$As$_{1-x-y}$Bi$_{x}$ alloys we employ simple cubic supercells containing 4096 atoms. We have previously demonstrated that this supercell size is sufficiently large to (i) mitigate finite size effects on the calculated electronic properties, by providing a suitably large basis of folded bulk states with which to describe N- and Bi-related localised states, and (ii) provide sufficient scope for the formation of a large variety of distinct local atomic environments to accurately reflect the short-range alloy disorder inherent in real materials. Our heterostructure calculations are performed for realistically sized, [001]-oriented GaN$_{y}$As$_{1-x-y}$Bi$_{x}$/GaAs QWs. The 24576-atom supercells used to study these structures have a total length of 24 nm along the [001] direction, and 4 nm along each of the [100] and [010] in-plane directions. The thickness of the GaN$_{y}$As$_{1-x-y}$Bi$_{x}$ QW layer is taken to be 8 nm in each case, with surrounding 8 nm thick GaAs barrier layers. The lateral extent and thickness of these QWs were chosen based on our analysis of GaAs$_{1-x}$Bi$_{x}$/GaAs QWs \cite{Usman_APL_2014} and MQWs \cite{Usman_PRMaterials_2018}, where we noted that (i) the lateral dimensions were sufficient to mitigate in-plane finite size effects, and (ii) the calculated properties were robust to QW thickness fluctuations $\sim \pm 1$ nm. All bulk and QW supercell calculations employ conventional Born-von Karman boundary conditions. In order to analyse the optical properties we compute the transverse electric (TE-) and transverse magnetic (TM-) polarised -- i.e.~polarised perpendicular or parallel to [001] -- optical transition strengths in the usual way, by using Fermi's golden rule in conjunction with the full supercell eigenstates (cf.~Eq.~(S4)).


\subsection{Continuum: 14-band \textbf{k}$\cdot$\textbf{p} Hamiltonian}
\label{sec:theory_kdotp}

Previous analysis has demonstrated it is useful to derive continuum models that describe perturbed band structure of GaN$_{y}$As$_{1-y}$ and GaAs$_{1-x}$Bi$_{x}$ alloys. Phenomenological approaches, principally the band-anticrossing (BAC) model, have originated from interpretation of spectroscopic data \cite{Shan_PRL_1999,Alberi_PRB_2007} and atomistic electronic structure calculations, \cite{Lindsay_PE_2004,Broderick_SST_2013,Bushell_submitted_2017} and are widely employed as a straightforward means by which to describe the evolution with alloy composition of the main features of the band structure of bulk materials and heterostructures. For GaN$_{y}$As$_{1-y}$ it is well established that the the CB structure can be described by a simple 2-band BAC model, in which the extended states of the GaAs host matrix CB edge interact with a set of higher energy N-related localised states. \cite{Shan_PRL_1999,Reilly_SST_2009} In Ga(In)N$_{y}$As$_{1-y}$ the composition dependence of the BAC interaction between these two sets of states results in a strong reduction of the alloy CB edge energy with increasing $y$. Similiar behaviour is present in GaAs$_{1-x}$Bi$_{x}$: the strong reduction (increase) and composition-dependent bowing of $E_{g}$ ($\Delta_{\scalebox{0.6}{\textrm{SO}}}$) can be described in terms of a valence band-anticrossing (VBAC) interaction, \cite{Alberi_PRB_2007,Broderick_SST_2013} between the extended states of the GaAs VB edge and localised impurity states associated with substitutional Bi impurities, which pushes the alloy VB edge upwards in energy with increasing $x$. While (V)BAC models generally omit effects associated with alloy disorder, they nonetheless provide reliable descriptions of the main features of the alloy band structure \cite{Reilly_SST_2002,Lindsay_SSE_2003,Broderick_SST_2013} and have been used as a basis to provide quantitative prediction of the properties of real dilute nitride and bismide photonic and photovoltaic devices. \cite{Hader_APL_2000,Marko_SR_2016}

We have previously demonstrated that an appropriate set of \textbf{k}$\cdot$\textbf{p} basis states for GaN$_{y}$As$_{1-x-y}$Bi$_{x}$ alloys must represent a minimum of 14 bands: the spin-degenerate CB, light-hole (LH), heavy-hole (HH) and SO bands of the GaAs host matrix (8 bands), the $A_{1}$-symmetric N localised states (of which there is one spin-degenerate set; 2 bands), and the $T_{2}$-symmetric Bi localised states (of which there are two spin-degenerate sets; 4 bands). Atomistic supercell calculations confirm that the respective impact of N and Bi on the band structure are decoupled in ordered GaN$_{y}$As$_{1-x-y}$Bi$_{x}$ alloys, confirming that an appropriate \textbf{k}$\cdot$\textbf{p} Hamiltonian can be constructed by directly superposing the separate VC, (V)BAC and strain-dependent contributions associated with N and Bi incorporation. \cite{Broderick_SST_2013} As such, the GaN$_{y}$As$_{1-x-y}$Bi$_{x}$ band structure then admits a simple interpretation in terms of the respective perturbation of the CB and VB separately by N- and Bi-related localised states, with minor VC changes to the CB (VB) structure due to Bi (N) incorporation. We parametrise the 14-band \textbf{k}$\cdot$\textbf{p} Hamiltonian directly from TB supercell calculations, obtaining the N- and Bi-related band parameters without the usual requirement to perform post hoc fitting to the results of alloy experimental data. \cite{Lindsay_PE_2004,Broderick_SST_2013}


\section{Results and Discussions}
\label{sec:results}


\subsection{Co-alloying N and Bi: dilute doping limit}
\label{sec:results_dilute}

Analysis of large, ordered $2M$-atom Ga$_{M}$N$_{1}$As$_{M-1}$ and Ga$_{M}$As$_{M-1}$Bi$_{1}$ supercells -- undertaken using the empirical pseudopotential \cite{Kent_PRL_2001,Kent_PRB_2001,Kent_SST_2002} and TB methods \cite{Lindsay_PB_2003,Lindsay_PE_2004,Usman_PRB_2011,Broderick_SST_2013} -- have revealed the mechanisms by which substitutional N and Bi impurities perturb the GaAs band structure. Specifically, it has been shown that the local relaxation of the crystal lattice (arising from differences in covalent radius) and charge transfer (arising from differences in atomic orbital energies) due to substitutional incorporation of an isolated N (Bi) impurity gives rise to highly localised states. These states are resonant with and couple to the GaAs CB (VB) in GaN$_{y}$As$_{1-y}$ (GaBi$_x$As$_{1-x}$), leading to hybridised alloy CB (VB) edge states containing an admixture of extended GaAs Bloch and localised N (Bi) character. These localised states can be described as linear combinations of the supercell eigenstates that fold back to $\Gamma$. Since more eigenstates fold back to $\Gamma$ with increasing supercell size, systematic analysis has demonstrated the need to use large supercells -- i.e.~ultra-dilute Bi and N compositions -- to accurately quantify the nature and impact of these localised impurity states in the dilute doping limit. \cite{Lindsay_PE_2004,Usman_PRB_2011,Broderick_SST_2013}


\begin{table*}[t!]
	\caption{\label{tab:band_edge_energies} Energies of the lowest energy CB ($E_{\protect\scalebox{0.6}{\textrm{CB}}}$), two highest energy VBs ($E_{\protect\scalebox{0.6}{\textrm{VB,1}}}$ and $E_{\protect\scalebox{0.6}{\textrm{VB,1}}}$), and SO band edge ($E_{\protect\scalebox{0.6}{\textrm{SO}}}$) calculated using the TB method for selected 4096-atom simple cubic GaN$_{y}$As$_{1-x-y}$Bi$_{x}$ supercells. The notation describing distinct (non-equivalent) local atomic configurations in the second column are defined in the text; accompanying schematic illustrations can be found in Fig.~S1 of the Supplementary Material. The spatial distance between the N and Bi atoms for each (unrelaxed) configuration is denoted by $r_{\protect\scalebox{0.6}{\textrm{N,Bi}}}$ and given in units of GaAs lattice constant $a_{0}$.}
	\begin{ruledtabular}
		\begin{tabular}{llccccc}
			Supercell & Configuration & $r_{\scalebox{0.6}{\textrm{N,Bi}}}$ ($a_{0}$) & $E_{\scalebox{0.6}{\textrm{CB}}}$ (eV) & $E_{\scalebox{0.6}{\textrm{V1}}}$ (eV) & $E_{\scalebox{0.6}{\textrm{V2}}}$ (eV) & $E_{\scalebox{0.6}{\textrm{SO}}}$ (eV) \\
			\hline
			Ga$_{2048}$As$_{2048}$                & GaAs                                   & ----- & $1.519$ &  $0.0010$ & $0.0010$ & $-0.3518$ \\
			\hline
			Ga$_{2048}$N$_{1}$As$_{2047}$         & GaAs:N$_{\scalebox{0.6}{\textrm{A}}}$  & ----- & $1.500$ &  $0.0013$ & $0.0011$ & $-0.3518$ \\
			Ga$_{2048}$As$_{2047}$Bi$_{1}$        & GaAs:Bi$_{\scalebox{0.6}{\textrm{B}}}$ & ----- & $1.518$ &  $0.0068$ & $0.0062$ & $-0.3521$ \\
			\hline
			Ga$_{2048}$N$_{1}$As$_{2046}$Bi$_{1}$ & GaAs:N$_{\scalebox{0.6}{\textrm{A}}}$Bi$_{\scalebox{0.6}{\textrm{B}}}$ & $\sqrt{3}$ & $1.500$ & $0.0152$ & $0.0131$ & $-0.3521$ \\
			Ga$_{2048}$N$_{1}$As$_{2046}$Bi$_{1}$ & GaAs:N$_{\scalebox{0.6}{\textrm{A}}}$Bi$_{\scalebox{0.6}{\textrm{C}}}$ & $\sqrt{2}$ & $1.506$ & $0.0154$ & $0.0127$ & $-0.3521$ \\
			Ga$_{2048}$N$_{1}$As$_{2046}$Bi$_{1}$ & GaAs:N$_{\scalebox{0.6}{\textrm{A}}}$Bi$_{\scalebox{0.6}{\textrm{D}}}$ & $\frac{1}{\sqrt{2}}$ & $1.495$ & $0.0196$ & $0.0115$ & $-0.3523$ \\
		\end{tabular}
	\end{ruledtabular}
\end{table*}

Here, we perform a similar analysis for a series of 4096-atom Ga$_{2048}$N$_{1}$As$_{2046}$Bi$_{1}$ supercells. We vary the relative positions of the N and Bi atoms to ascertain (i) any changes to their respective impact on the electronic structure, and (ii) the scale of any interactions between N- and Bi-related localised states. In each case we compute the fractional GaAs $\Gamma$ (Bloch) character \cite{Reilly_JPCM_2004,Usman_PRB_2013} of the supercell zone-centre eigenstates in order to quantify the changes in the electronic structure associated with the formation of different N and Bi local atomic environments. Full details of the calculated $\Gamma$ character spectra are presented in Sec.~S2 of the Supplementary Material.

Table~\ref{tab:band_edge_energies} summarises the results of our calculations for a range of Ga$_{2048}$N$_{1}$As$_{2046}$Bi$_{1}$ supercells. We use subscripts B, C and D to describe the positions of Bi atoms relative to a N atom at position A. Prior to relaxation of the atomic positions, the vectors describing the separation between these atomic sites are: $\textbf{r}_{\scalebox{0.6}{\textrm{AB}}} = a_{0} ( \widehat{x} + \widehat{y} + \widehat{z} )$, $\textbf{r}_{\scalebox{0.6}{\textrm{AC}}} = a_{0} ( \widehat{x} + \widehat{y} )$ and $\textbf{r}_{\scalebox{0.6}{\textrm{AD}}} = \frac{ a_{0} }{ 2 } ( \widehat{x} + \widehat{y} )$, where $a_{0}$ is the GaAs lattice constant. For example, the notation GaAs:N$_{\scalebox{0.6}{\textrm{A}}}$Bi$_{\scalebox{0.6}{\textrm{B}}}$ describes a supercell in which the Bi atom is oriented along [111] relative to the N atom, with the N and Bi atoms occupying the same position in a given pair of neighbouring 8-atom simple cubic unit cells, while the notation GaAs:N$_{\scalebox{0.6}{\textrm{A}}}$Bi$_{\scalebox{0.6}{\textrm{D}}}$ describes a [110] oriented N-Ga-Bi complex in which the N and Bi atoms share a common Ga nearest neighbour.

We begin with a Ga$_{2048}$N$_{1}$As$_{2046}$Bi$_{1}$ supercell in which the N and Bi atoms are separated by $\sqrt{3}a_0$ (where they do not form a pair or cluster) and note that the associated reduction in symmetry, arising from relaxation of the crystal lattice about the impurity sites, lifts the degeneracy of the predominantly GaAs LH- and HH-like VB edge states. As the N and Bi atoms are brought closer together the splitting between these states increases due to the resultant larger local relaxation of the lattice, with the largest calculated splitting of $\approx 10$ meV in a Ga$_{2048}$N$_{1}$As$_{2046}$Bi$_{1}$ supercell occurring for an N-Ga-Bi complex oriented along the [110] direction (in which the N and Bi atoms are second-nearest neighbours and share a Ga nearest neighbour). We note that this lifting of the VB edge degeneracy due to a reduction in symmetry is consistent with that calculated previously for disordered GaAs$_{1-x}$Bi$_{x}$. \cite{Usman_PRB_2011,Usman_PRB_2013,Bannow_arxiv_2017} In the Ga$_{2048}$N$_{1}$As$_{2047}$ and Ga$_{2048}$As$_{2047}$Bi$_{1}$ supercells we note the presence of a small ($< 1$ meV) splitting of the LH- and HH-like VB edge states. We do not expect a lifting of the VB edge degeneracy in these ordered structures: the observed splitting is due to small residual strains associated with the convergence of the relaxation of the atomic positions using the valence force field model. \cite{Usman_PRB_2011,Usman_PRB_2013}

From Table~\ref{tab:band_edge_energies}, we see that as the N and Bi atoms are brought closer together the splitting between the two highest energy VB states ($E_{v1}$ and $E_{v2}$) increases due to the resultant larger local relaxation of the lattice. The largest calculated splitting of $\approx 10$ meV in a Ga$_{2048}$N$_{1}$As$_{2046}$Bi$_{1}$ supercell occurs when the N and Bi atoms are second-nearest neighbours sharing a Ga nearest neighbour (in the GaAs:N$_{\scalebox{0.6}{\textrm{A}}}$Bi$_{\scalebox{0.6}{\textrm{D}}}$ supercell). We note that this lifting of the VB edge degeneracy due to a reduction in symmetry is consistent with that calculated previously for disordered GaAs$_{1-x}$Bi$_{x}$. \cite{Usman_PRB_2011,Usman_PRB_2013,Bannow_arxiv_2017} We note also that the calculated shift in the VB edge energy in the supercells containing both N and Bi is significantly larger than that in the equivalent N-free supercells. Our analysis suggests that this is a result of the large local relaxation of the lattice about the N atomic site which, within the framework of our TB model, generates significant shifts in the energies of the $p$ orbitals localised on the N atom. However, we calculate that this trend does not persist beyond the ultra-dilute regime, with the calculated overall shift in the VB edge energy with composition in GaN$_{y}$As$_{1-x-y}$Bi$_{x}$ closely tracking that in GaAs$_{1-x}$Bi$_{x}$ (cf.~Sec.~\ref{sec:results_disordered}).

The calculated trends in the impact of interaction between N- and Bi-related localised states is consistent with previous calculations, \cite{Broderick_SST_2013} and can be understood generally on the basis of their respective independent impact in Ga$_{2048}$N$_{1}$As$_{2047}$ and Ga$_{2048}$As$_{2047}$Bi$_{1}$ supercells. For the CB edge state, having energy $E_{\scalebox{0.6}{\textrm{CB}}}$, the impact of the interaction between the N and Bi atoms is minimal: the overall character of the CB edge is predominantly determined by the impact of the localised resonant state associated with the N atom. Even in the GaAs:N$_{\scalebox{0.6}{\textrm{A}}}$Bi$_{\scalebox{0.6}{\textrm{B}}}$ supercell -- in which the N and Bi atoms are closest, and hence the interaction of their associated localised states maximised -- the largest calculated Bi-induced shift in $E_{\scalebox{0.6}{\textrm{CB}}}$ is minimal when compared to the overall shift in $E_{\scalebox{0.6}{\textrm{CB}}}$ calculated in an equivalent Bi-free supercell. In this case the character of the calculated GaN$_{y}$As$_{1-x-y}$Bi$_{x}$ CB edge state results from a combination of an N-induced hybridisation and reduction in energy (described by the conventional 2-band BAC model), in addition to small further changes associated with (i) the VC-like Bi-induced reduction in $E_{\scalebox{0.6}{\textrm{CB}}}$, and (ii) the local compressive strain associated with lattice relaxation about the Bi atomic site. Contrary to the trends observed for the CB and VB edges, we note that the SO band edge energy is relatively unaffected by N incorporation: the calculated trends in the energy of $\Gamma$ character of the SO band edge states are essentially identical to those in GaAs$_{1-x}$Bi$_{x}$, with the impact of N manifesting primarily via small energy shifts associated with local lattice relaxation.

Next, we turn our attention to the localised states associated with N and Bi, which we construct explicitly for each supercell. \cite{Lindsay_PE_2004,Reilly_JPCM_2004,Usman_PRB_2011} Our analysis reveals a somewhat surprising feature: the overall nature and character of the N- (Bi-) related localised states are found to be effectively identical to those in an equivalent Bi-free Ga$_{2048}$N$_{1}$As$_{2047}$ (N-free Ga$_{2048}$As$_{2047}$Bi$_{1}$) supercell. In all cases we find that the character of the band edge eigenstates is largely retained: the CB (VB) edge eigenstates are a linear combination of the unperturbed GaAs CB (VB) edge and N (Bi) localised states, describable via the same 2- (4-) band (V)BAC model as in GaN$_{y}$As$_{1-y}$ (GaAs$_{1-x}$Bi$_{x}$). \cite{Lindsay_PE_2004,Usman_PRB_2011} This confirms that the impact of N (Bi) on the CB (VB) structure is effectively independent of co-alloying with Bi (N) and, as we will see below, that general trends in the evolution of the GaN$_{y}$As$_{1-x-y}$Bi$_{x}$ band structure can be described to a reasonable degree of accuracy by superposing the established description of impact of both N and Bi incorporation. \cite{Broderick_SST_2013}


\begin{figure*}[t!]
	\includegraphics[width=1.00\textwidth]{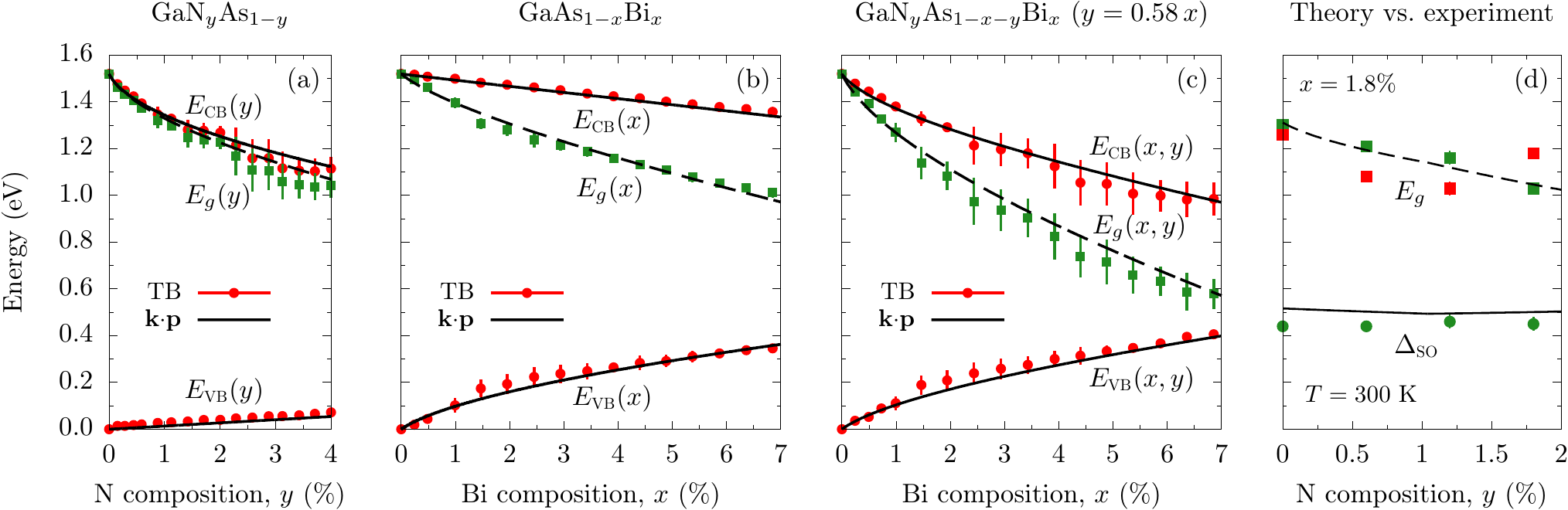}
 	\caption{(a), (b) and (c) Calculated variation of the CB and VB edge energies $E_{\protect\scalebox{0.6}{\textrm{CB}}}$ and $E_{\protect\scalebox{0.6}{\textrm{VB}}}$, and band gap $E_{g} = E_{\protect\scalebox{0.6}{\textrm{CB}}} - E_{\protect\scalebox{0.6}{\textrm{VB}}}$, with alloy composition in GaN$_{y}$As$_{1-y}$, GaAs$_{1-x}$Bi$_{x}$ and GaN$_{y}$As$_{1-x-y}$Bi$_{x}$ (lattice-matched to GaAs). Closed red circles (green squares) show the values of $E_{\protect\scalebox{0.6}{\textrm{CB}}}$ and $E_{\protect\scalebox{0.6}{\textrm{VB}}}$ ($E_{g}$) averaged at each composition over those calculated for a series of free-standing, disordered 4096-atom supercells using the $sp^{3}s^{*}$ tight-binding model. Error bars denote the standard deviation about these average values, computed from the corresponding energies for the distinct supercells considered at each composition. Solid (dashed) black lines show $E_{\protect\scalebox{0.6}{\textrm{CB}}}$ and $E_{\protect\scalebox{0.6}{\textrm{VB}}}$ ($E_{g}$) calculated using the 14-band \textbf{k}$\cdot$\textbf{p} Hamiltonian. (d) Variation of $E_{g}$ and $\Delta_{\protect\scalebox{0.6}{SO}}$ with N composition $y$ in GaN$_{y}$As$_{0.982-y}$Bi$_{0.018}$ ($x = 1.8$\%) calculated using the 14-band \textbf{k}$\cdot$\textbf{p} Hamiltonian (dashed and solid black lines, respectively), compared to the results of room temperature spectroscopic ellipsometry (SE; closed green squares and circles) and photo-modulated reflectance (PR; closed red squares) measurements. The experimental data are from Ref.~\onlinecite{Bushell_JAP_2018}.}
	\label{fig:band_edge_energies}
 \end{figure*}


\subsection{Band edge energies in disordered alloys}
\label{sec:results_disordered}

Having analysed the impact of co-alloying N and Bi in the dilute doping limit, we now turn our attention to the evolution of the electronic structure with alloy composition in disordered alloys. We restrict our attention to lattice-matched, 4096-atom supercells in which the N and Bi compositions have been chosen to produce net zero macroscopic strain with respect to GaAs -- i.e.~we require $y = 0.58 \, x$ so that the GaN$_{y}$As$_{1-x-y}$Bi$_{x}$ lattice constant computed using V\'{e}gard's law is equal to that of GaAs. At each alloy composition we compute the electronic structure of five distinct supercells in which the N and Bi atoms are substituted at randomly chosen sites on the anion sublattice. To determine the composition dependence of the band edge energies we average over the results of these five distinct calculations at each composition. As a reference for the GaN$_{y}$As$_{1-x-y}$Bi$_{x}$ calculations, we have performed the same analysis for equivalent Bi- (N-) free GaN$_{y}$As$_{1-y}$ (GaAs$_{1-x}$Bi$_{x}$) supercells. The results of these calculations are summarised in Figs.~\ref{fig:band_edge_energies}(a), \ref{fig:band_edge_energies}(b) and \ref{fig:band_edge_energies}(c) for GaN$_{y}$As$_{1-y}$, GaAs$_{1-x}$Bi$_{x}$ and GaN$_{y}$As$_{1-x-y}$Bi$_{x}$, respectively.

Beginning with GaN$_{y}$As$_{1-y}$, the closed red circles in Fig.~\ref{fig:band_edge_energies}(a) show the calculated dependence of the CB and VB edge energies $E_{\scalebox{0.6}{\textrm{CB}}}$ and $E_{\scalebox{0.6}{\textrm{VB}}}$ on $y$. Our calculations reproduce the well-known trends for GaN$_{y}$As$_{1-y}$. Firstly, N incorporation causes a rapid decrease and large composition dependent bowing of $E_{\scalebox{0.6}{\textrm{CB}}}$, with the calculated CB edge states consisting of an admixture of GaAs CB edge Bloch and N localised state character. \cite{Lindsay_SSC_2001,Lindsay_PE_2004} Secondly, the VB and SO edge energies are relatively unperturbed from those calculated in ordered alloy supercells, and are well described via conventional VC energy shifts as $E_{\scalebox{0.6}{\textrm{VB}}} (y) = E_{\scalebox{0.6}{\textrm{VB}}} (0) + \kappa_{\scalebox{0.6}{\textrm{N}}} \, y$ and $E_{\scalebox{0.6}{\textrm{SO}}} (y) = E_{\scalebox{0.6}{\textrm{SO}}} (0) - \gamma_{\scalebox{0.6}{\textrm{N}}} \, y$, where $E_{\scalebox{0.6}{\textrm{VB}}} (0)$ and $E_{\scalebox{0.6}{\textrm{SO}}} (0)$ are the corresponding unperturbed GaAs band edge energies. \cite{Broderick_SST_2013} Thirdly, the band gap $E_{g} = E_{\scalebox{0.6}{\textrm{CB}}} - E_{\scalebox{0.6}{\textrm{VB}}}$ (depicted by closed green squares in Fig.~\ref{fig:band_edge_energies}(a)) is calculated to decrease by $\approx 180$ meV when 1\% of the As atoms are replaced by N, in good agreement with a range of experimental measurements. \cite{Reilly_SST_2009}

The closed red circles in Fig.~\ref{fig:band_edge_energies}(b) show the calculated dependence of $E_{\scalebox{0.6}{\textrm{CB}}}$ and $E_{\scalebox{0.6}{\textrm{VB}}}$ on $x$ in GaAs$_{1-x}$Bi$_{x}$. Again we recover the experimentally observed trends. Bi incorporation causes a rapid decrease of the band gap (closed green squares), which is characterised by strong composition dependent bowing and is qualitatively similar to that in GaN$_{y}$As$_{1-y}$, but instead originates from strong upward bowing of $E_{\scalebox{0.6}{\textrm{VB}}}$. \cite{Alberi_PRB_2007} The calculated VB edge eigenstates consist of an admixture of GaAs VB Bloch and Bi localised state character, \cite{Usman_PRB_2011,Broderick_SST_2013} and the calculated $\approx 100$ meV decrease in $E_{g}$ in going from GaAs to GaAs$_{0.99}$Bi$_{0.01}$ is in good agreement with experimental measurements. \cite{Usman_PRB_2011,Batool_JAP_2012} We note that our calculations correctly describe that the decrease of $E_{g}$ due to N incorporation is larger than that associated with Bi incorporation at fixed composition, reflecting the larger differences in covalent radius and electronegativity between N and As than between As and Bi. We further find that $E_{\scalebox{0.6}{\textrm{CB}}}$ and the SO band edge energy $E_{\scalebox{0.6}{\textrm{SO}}}$ in GaAs$_{1-x}$Bi$_{x}$ are well described via conventional VC energy shifts as $E_{\scalebox{0.6}{\textrm{CB}}} (x) = E_{\scalebox{0.6}{\textrm{VB}}} (0) - \alpha_{\scalebox{0.6}{\textrm{Bi}}} \, x$ and $E_{\scalebox{0.6}{\textrm{SO}}} (x) = E_{\scalebox{0.6}{\textrm{SO}}} (0) - \gamma_{\scalebox{0.6}{\textrm{Bi}}} \, x$. \cite{Broderick_SST_2013} The calculated VC parameters $\kappa_{\scalebox{0.6}{\textrm{N,Bi}}}$, $\alpha_{\scalebox{0.6}{\textrm{N,Bi}}}$ and $\gamma_{\scalebox{0.6}{\textrm{N,Bi}}}$ are given, along with the remainder of the parameters of the 14-band \textbf{k}$\cdot$\textbf{p} model, in Sec.~S1.B of the Supplementary Material.

The error bars in Figs.~\ref{fig:band_edge_energies}(a) and~\ref{fig:band_edge_energies}(b) denote the standard deviations of the corresponding energies, computed from the values of $E_{\scalebox{0.6}{\textrm{CB}}}$, $E_{\scalebox{0.6}{\textrm{VB}}}$ and $E_{g}$ calculated for each of the distinct supercells considered at fixed alloy composition. For GaN$_{y}$As$_{1-y}$ we note that the calculated standard deviations for $E_{\scalebox{0.6}{\textrm{CB}}}$ (and hence $E_{g}$) increase strongly with increasing $y$, from 6 meV at an ultra-dilute N composition $y = 0.15$\% to $> 80$ meV for $2\% \lesssim y \lesssim 3$\%. This reflects that N-related alloy disorder has a strong impact on the GaN$_{y}$As$_{1-y}$ CB structure, with the different spatially random distributions of N atoms in the supercells considered leading to large differences in the computed CB edge energy at fixed $y$. Conversely, we compute negligible standard deviations $\lesssim 3$ meV for $E_{\scalebox{0.6}{\textrm{VB}}}$ across the range of $y$ considered, reflecting the weak impact of N incorporation on the VB structure. We find that these trends are reversed in GaAs$_{1-x}$Bi$_{x}$ alloys, where we compute negligible standard deviations $\lesssim 3$ meV for $E_{\scalebox{0.6}{\textrm{CB}}}$ at each Bi composition $x$, reflecting that VC contributions dominate the CB and SO band edge character. However, we compute larger standard deviations for $E_{\scalebox{0.6}{\textrm{VB}}}$, reflecting the important role played by Bi-related localised states and alloy disorder in determining the nature of the GaAs$_{1-x}$Bi$_{x}$ VB edge states. However, the maximum calculated value of 43 meV for the standard deviation of $E_{\scalebox{0.6}{\textrm{VB}}}$ at $x = 1.95$\% is approximately one-half of that calculated for GaN$_{y}$As$_{1-y}$ at similar composition indicating, as expected, that N-related alloy disorder more strongly perturbs the electronic structure.

Figure~\ref{fig:band_edge_energies}(c) summarises the results of the disordered alloy supercell calculations for GaN$_{y}$As$_{1-x-y}$Bi$_{x}$ lattice-matched to GaAs. The variation of $E_{\scalebox{0.6}{\textrm{CB}}}$, $E_{\scalebox{0.6}{\textrm{VB}}}$ and $E_{g}$ are shown here as a function of the Bi composition $x$, for N compositions $y = 0.58 \, x$. The calculated reduction of $E_{g}$ -- approximately 220 meV (400 meV) in a lattice-matched alloy having $x = 1$\% ($y = 1$\%) -- is significantly larger than that in either GaN$_{y}$As$_{1-y}$ or GaAs$_{1-x}$Bi$_{x}$. That this giant band gap bowing can be achieved in alloys which are lattice-matched to GaAs suggests significant potential for applications at infrared wavelengths. We note (i) the decrease in $E_{\scalebox{0.6}{\textrm{CB}}}$ is comparable to, but slightly larger than, that in GaN$_{y}$As$_{1-y}$, and (ii) the increase in $E_{\scalebox{0.6}{\textrm{VB}}}$ is approximately equal to that in GaAs$_{1-x}$Bi$_{x}$. $\Delta_{\scalebox{0.6}{\textrm{SO}}}$ is calculated to increase by approximately 70 meV per \% Bi in lattice-matched GaN$_{y}$As$_{1-x-y}$Bi$_{x}$ -- i.e.~by approximately the same amount as in GaAs$_{1-x}$Bi$_{x}$, reflecting the weak impact of N incorporation on the VB structure.

As in Figs.~\ref{fig:band_edge_energies}(a) and~\ref{fig:band_edge_energies}(b), the error bars in Fig.~\ref{fig:band_edge_energies}(c) denote the standard deviations computed for $E_{\scalebox{0.6}{\textrm{CB}}}$, $E_{\scalebox{0.6}{\textrm{VB}}}$ and $E_{g}$ using the corresponding calculated values for the distinct supercells considered at each fixed alloy composition. As expected on the basis of the trends discussed above for ternary GaN$_{y}$As$_{1-y}$ and GaAs$_{1-x}$Bi$_{x}$ alloys, (i) the impact of N and Bi incorporation leads to large respective computed standard deviations in $E_{\scalebox{0.6}{\textrm{CB}}}$ and $E_{\scalebox{0.6}{\textrm{VB}}}$, and (ii) the magnitude of the computed standard deviation for $E_{\scalebox{0.6}{\textrm{CB}}}$ at fixed composition is larger than that associated with $E_{\scalebox{0.6}{\textrm{VB}}}$. We note then that the standard deviation associated with $E_{g}$, the maximum calculated value of which was 98 meV at $x = 3.9$\%, is typically larger in GaN$_{y}$As$_{1-x-y}$Bi$_{x}$ than that in either GaN$_{y}$As$_{1-y}$ or GaAs$_{1-x}$Bi$_{x}$ having the same N or Bi composition, but is in general broadly comparable in magnitude to that associated with GaN$_{y}$As$_{1-y}$. Overall, these results reaffirm that N- and Bi-related alloy disorder leads to significant inhomogeneous spectral broadening of the band edge transitions in GaN$_{y}$As$_{1-y}$ and GaAs$_{1-x}$Bi$_{x}$ alloys, and demonstrates that while alloy disorder effects can be expected to have an important impact on the GaN$_{y}$As$_{1-x-y}$Bi$_{x}$ electronic structure, the magnitude of such effects should somewhat exceed but nonetheless be broadly comparable to those in GaN$_{y}$As$_{1-y}$.

To confirm that these trends can be described as a direct combination of the separate impact of N (Bi) incorporation primarily on the CB (VB) structure, we have also used the 14-band \textbf{k}$\cdot$\textbf{p} model to calculate the composition dependent band edge energies. The results of the \textbf{k}$\cdot$\textbf{p} calculations of $E_{\scalebox{0.6}{\textrm{CB}}}$ and $E_{\scalebox{0.6}{\textrm{VB}}}$ (solid black lines) and $E_{g}$ (dashed black lines) for GaN$_{y}$As$_{1-y}$, GaAs$_{1-x}$Bi$_{x}$ and GaN$_{y}$As$_{1-x-y}$Bi$_{x}$ are shown respectively in Figs.~\ref{fig:band_edge_energies}(a), 1(b) and 1(c). In GaN$_{y}$As$_{1-y}$ (GaAs$_{1-x}$Bi$_{x}$) this corresponds to a 10- (12-) band \textbf{k}$\cdot$\textbf{p} model of the band structure. In all cases we note that the variation of $E_{\scalebox{0.6}{\textrm{CB}}}$ and $E_{\scalebox{0.6}{\textrm{VB}}}$ with composition are in good overall agreement with the results of the disordered supercell calculations. We note some minor deviation between the calculated variation of $E_{g}$ in GaN$_{y}$As$_{1-y}$ and GaN$_{y}$As$_{1-x-y}$Bi$_{x}$ using the TB and \textbf{k}$\cdot$\textbf{p} models, which we identify as being associated with the relatively stronger impact of N-related cluster states -- neglected in the \textbf{k}$\cdot$\textbf{p} model -- on the CB edge. \cite{Reilly_JPCM_2004,Reilly_SST_2009,Broderick_nitride_chapter_2017} These results suggest overall that the evolution of the main features of the GaN$_{y}$As$_{1-x-y}$Bi$_{x}$ band structure (i) is primarily determined by the influence of localised states associated with independent N and Bi impurities, and (ii) can be reliably described using an extended \textbf{k}$\cdot$\textbf{p} Hamiltonian whose basis explicitly includes these localised states and their coupling to the GaAs host matrix band edge states.


\begin{table*}[t!]
	\caption{\label{tab:qw_structures} Details of the GaN$_{y}$As$_{1-x-y}$Bi$_{x}$/GaAs structures analysed in Secs. 3.3 and 3.4. Structures 1 and 2 contain ternary (N-free) GaAs$_{1-x}$Bi$_{x}$/GaAs QWs, while structures 3 and 4 contain quaternary GaN$_{y}$As$_{1-x-y}$Bi$_{x}$ QWs. In addition to the Bi and N compositions $x$ and $y$ we summarise the results of the 14-band \textbf{k}$\cdot$\textbf{p} calculations, which are independent of alloy disorder at fixed composition. We provide the computed in-plane component $\epsilon_{xx}$ ($= \epsilon_{yy}$, compressive in all structures) of the macroscopic strain, as well as the CB and VB band offsets ($\Delta E_{\protect\scalebox{0.6}{\textrm{CB}}}$ and $\Delta E_{\protect\scalebox{0.6}{\textrm{HH}}}$), ground state transition energy $e1$-$h1$ at $T = 300$ K, and the corresponding ground state emission/absorption wavelength for the $e1$-$h1$ transition.}
	\begin{ruledtabular}
		\begin{tabular}{cccccccc}
			Structure & $x$ (\%) & $y$ (\%) & $\epsilon_{xx}$ (\%) & $\Delta E_{\scalebox{0.6}{\textrm{CB}}}$ (meV) & $\Delta E_{\scalebox{0.6}{\textrm{HH}}}$ (meV) & $e1$-$h1$ (eV) & $e1$-$h1$ (nm) \\
			\hline
			1 & 6.25 & ----- & $-0.74$ & 106 & 359 & 1.097 & 1130 \\
			2 & 9.00 & ----- & $-1.06$ & 153 & 456 & 0.960 & 1292 \\
			\hline
			3 & 6.25 & 2.50  & $-0.23$ & 419 & 366 & 0.791 & 1568 \\
			4 & 9.00 & 1.00  & $-0.86$ & 306 & 459 & 0.808 & 1535 \\
		\end{tabular}
	\end{ruledtabular}
\end{table*}

Finally, Fig.~\ref{fig:band_edge_energies}(d) compares the variation of $E_{g}$ (solid black lines) and $\Delta_{\scalebox{0.6}{\textrm{SO}}}$ (dashed black lines) with N composition $y$, calculated using the 14-band \textbf{k}$\cdot$\textbf{p} model, to photo-modulated reflectance (PR) and spectroscopic ellipsometry (SE) measurements undertaken on a series of pseudomorphically strained GaN$_{y}$As$_{0.982-y}$Bi$_{0.018}$ ($x = 1.8$\%) samples grown on GaAs via metal-organic vapour phase epitaxy. \cite{Bushell_JAP_2018} The calculated variation of $E_{g}$ and $\Delta_{\scalebox{0.6}{\textrm{SO}}}$ with $y$ are in good overall agreement with experiment. Firstly, the 14-band model accurately describes the measured large reduction in $E_{g}$ compared to that in GaAs, and captures the evolution of $E_{g}$ with increasing N composition. Secondly, the 14-band model describes well the overall magnitude of $\Delta_{\scalebox{0.6}{\textrm{SO}}}$ -- which is larger than that in GaAs due to the presence of Bi -- and that incorporating N tends to have little impact on $\Delta_{\scalebox{0.6}{\textrm{SO}}}$, with the measured and calculated values remaining approximately constant across the range of N compositions considered, confirming the predicted weak impact of N incorporation on the VB structure.


\subsection{Electronic properties of GaN$_{y}$As$_{1-x-y}$Bi$_{x}$/GaAs quantum wells}
\label{sec:results_qw_electronic}

In order to realise photonic devices based on GaN$_{y}$As$_{1-x-y}$Bi$_{x}$ alloys, in practice it will likely be required to develop quantum-confined heterostructures. We therefore elucidate and analyse general features of the electronic and optical properties of GaN$_{y}$As$_{1-x-y}$Bi$_{x}$/GaAs QWs, focusing in particular on the bound lowest energy electron and highest energy hole states $e1$ and $h1$. In order to account for alloy disorder effects, for each QW structure having fixed N and Bi composition we consider ten distinct supercells having different statistically random spatial distributions (RDs) of substitutional N and Bi atoms at anion lattice sites in the GaN$_{y}$As$_{1-x-y}$Bi$_{x}$ QW layer. We compare the results of atomistic TB calculations to those obtained for the same structures using the continuum 14-band \textbf{k}$\cdot$\textbf{p} model in the envelope function approximation (EFA), in which the QWs are treated as idealised one-dimensional structures. Since the 14-band model does not explicitly account for the presence of alloy disorder, the results of the \textbf{k}$\cdot$\textbf{p} calculations provide a reference against which to highlight the role played by alloy disorder in the full atomistic calculations.

Four compressively strained QW structures are considered: structures 1 and 2 are N-free GaAs$_{1-x}$Bi$_{x}$/GaAs QWs having respective Bi compositions $x = 6.25$ and $9$\%, while structures 3 and 4 are GaN$_{y}$As$_{1-x-y}$Bi$_{x}$/GaAs QWs having respective Bi and N compositions $x = 6.25$\%, $y = 2.5$\% and $x = 9$\%, $y = 1$\%. These structures are described in Table~\ref{tab:qw_structures}, and the simulated geometries are as described in Sec.~\ref{sec:theory_tb} above. The N and Bi compositions for structures 3 and 4 were chosen to produce ground state $e1$-$h1$ transition energies close to 0.8 eV, so as to analyse the QW properties in the composition ranges of interest for applications in the technologically important 1.55 $\mu$m wavelength range. Table~\ref{tab:qw_structures} also summarises the results of the 14-band \textbf{k}$\cdot$\textbf{p} calculations for each QW structure, including the in-plane compressive strain $\epsilon_{xx}$, CB and VB offsets $\Delta E_{\scalebox{0.6}{\textrm{CB}}}$ and $\Delta E_{\scalebox{0.6}{\textrm{HH}}}$, and QW band gaps ($e1$-$h1$ transition energies). Based on our analysis in Sec.~\ref{sec:results_disordered} we conclude that GaN$_{y}$As$_{1-x-y}$Bi$_{x}$/GaAs heterostructures have large type-I band offsets and can therefore be expected to possess intrinsically high electron-hole spatial overlap, suggesting the possibility to achieve good optical efficiency and indicating potential for the development of light-emitting/absorbing devices. We calculate respective CB and VB offsets $\Delta E_{\scalebox{0.6}{\textrm{CB}}} = 106$ meV and $\Delta E_{\scalebox{0.6}{\textrm{HH}}} = 359$ meV in structure 1 ($x = 6.25$\%), compared with 153 and 456 meV in structure 2 ($x = 9$\%). In these N-free structures Bi incorporation brings about compressive strain and large VB offsets $\Delta E_{\scalebox{0.6}{\textrm{HH}}}$, and while $\Delta E_{\scalebox{0.6}{\textrm{CB}}}$ is considerably smaller it is nonetheless sufficiently large to provide good confinement of electrons and holes to facilitate efficient photon emission/absorption. \cite{Broderick_IEEEJSTQE_2015}

Beginning with structure 1 and incorporating 2.5\% N in the QW layer to obtain structure 3 increases the ground state transition wavelength by $\approx 450$ nm to 1.55 $\mu$m, while simultaneously reducing the net compressive strain by a factor of approximately three to $\epsilon_{xx} = -0.23$\%. Comparing the calculated band offsets for structures 1 and 3 we observe that N incorporation significantly increases $\Delta E_{\scalebox{0.6}{\textrm{CB}}}$, by a factor of approximately four to 419 meV, corresponding to the strong N-induced reduction of the QW CB edge energy. The VB offset is effectively unchanged, again reflecting that N incorporation has little effect on the VB. We observe a similar trend in incorporating 1\% N to go from structure 2 to structure 4. In all cases we find that the trends in the net macroscopic strain, QW band offsets, and band gap calculated using the 14-band model are consistent with those obtained from the full atomistic calculations.


\begin{figure*}[t!]
	\includegraphics[width=1.00\textwidth]{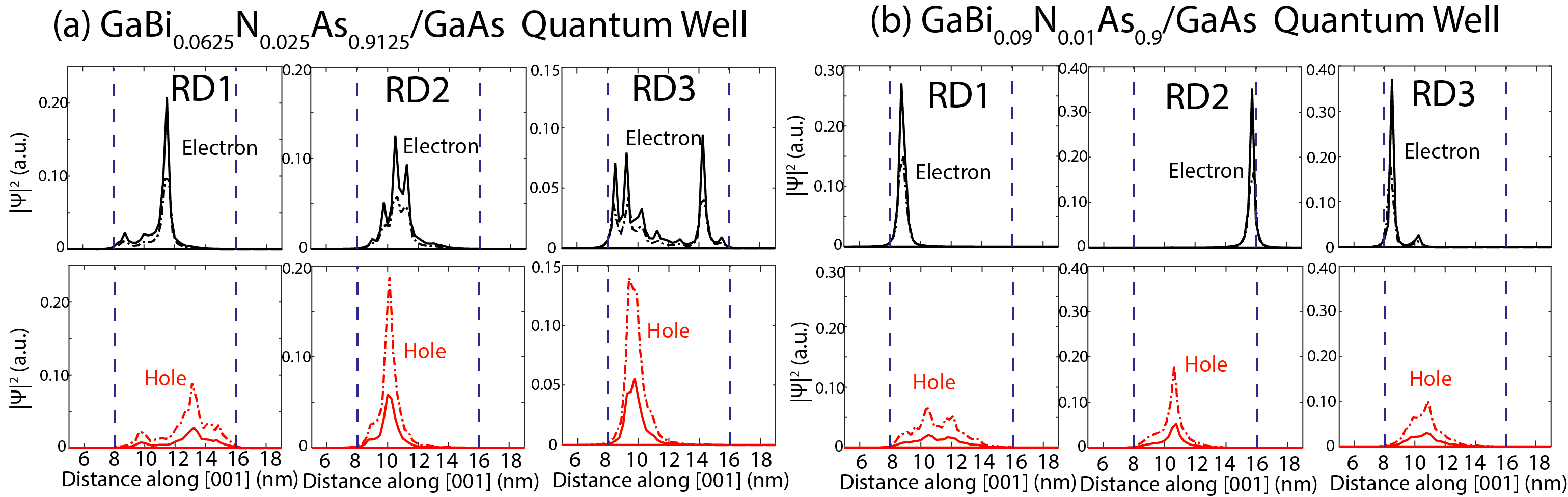}
	\caption{Probability density associated with the lowest energy CB electron state ($e1$; upper row) and highest energy VB hole state ($h1$; lower row) in 8 nm thick (a) GaN$_{0.025}$As$_{0.9125}$Bi$_{0.0625}$/GaAs ($x = 6.25$\%, $y = 2.5$\%), and (b) GaN$_{0.01}$As$_{0.90}$Bi$_{0.09}$/GaAs ($x = 9$\%, $y = 1$\%) QWs. Solid (dash-dotted) black lines and solid (dash-dotted) red lines respectively denote the $e1$ and $h1$ probability densities at cation (anion) sites calculated using the TB model, obtained at each position $z$ along the [001] growth direction by summing over the probability densities associated with each atom in the plane. Dashed blue lines denote the well/barrier interfaces. The TB calculations were performed for ten supercells containing different random spatial distributions (RDs) of substitutional N and Bi atoms on the anion sublattice: results are presented here for three representative structures in each case, with plots for additional GaN$_{y}$As$_{1-x-y}$Bi$_{x}$/GaAs structures, as well as for GaAs$_{1-x}$Bi$_{x}$/GaAs structures, provided in Figs.~S5 and S6 of the Supplementary Material.}
  	\label{fig:qw_probability_density}
\end{figure*}

Figures~\ref{fig:qw_probability_density}(a) and 2(b) show the probability density for the $e1$ (upper panels; black lines) and $h1$ (lower panels; red lines) states in the GaN$_{y}$As$_{1-x-y}$Bi$_{x}$/GaAs QWs (a) structure 3, and (b) structure 4, calculated using the TB method. For each structure we have plotted probability densities for three of the ten different RDs of Bi and N atoms considered; plots for additional RDs can be found in Fig.~S6 of the Supplementary Material. Solid and dashed lines respectively denote the probability density projected to cations and anions, calculated using the TB method at each fixed position $z$ along [001] by summing over the probability density associated with all atoms in the plane perpendicular to [001]. Firstly, we note that the calculated probability densities in the N-free QWs, structures 1 and 2 -- which, for brevity, are provided in Sec.~S3 of the Supplementary Material -- demonstrate that short-range alloy disorder strongly perturbs the hole states in GaAs$_{1-x}$Bi$_{x}$/GaAs QWs, while the electron states in the relatively unperturbed CB can be well described using the conventional EFA. The GaN$_{y}$As$_{1-x-y}$Bi$_{x}$/GaAs $h1$ probability densities calculated using the TB method are qualitatively the same as those in the N-free structures, as expected given the weak impact of N incorporation on the VB structure. The $h1$ eigenstates depart from a conventional envelope function-like behaviour: we calculate that, in a given structure, $h1$ tends to localise preferentially within the QW about regions of locally high Bi composition -- i.e.~about pairs and larger clusters of Bi atoms. In the CB we note that introducing N brings about a marked qualitative change in the nature of the $e1$ eigenstates: the $e1$ probability density in the presence of N mirrors that of $h1$ in the presence of Bi, reflecting that N localised states strongly perturb the CB structure. As we have identified in our previous analysis of the GaN$_{y}$As$_{1-y}$ (GaAs$_{1-x}$Bi$_{x}$) electronic structure, the observed strong localisation of the CB (VB) edge eigenstates generally reflects strong hybridisation of the extended GaAs band edge states with a multiplicity of N (Bi) localised states associated with pairs and larger clusters of substitutional N (Bi) impurities, resulting in significant degradation of the Bloch character of the corresponding alloy eigenstates. \cite{Reilly_JPCM_2004,Usman_PRB_2011,Usman_PRB_2013} In the GaN$_{y}$As$_{1-x-y}$Bi$_{x}$ QW structures considered here this behaviour is reflected in the near complete breakdown of the EFA description of the carrier probability densities due to strong electron and hole localisation, with the details of the QW electronic properties being strongly dependent on the precise nature of the short-range alloy disorder present in the GaN$_{y}$As$_{1-x-y}$Bi$_{x}$ layer of the structure.

Figures~\ref{fig:qw_probability_density_avg}(a) -- 3(d) provide a summary of the calculated $e1$ and $h1$ probability densities for structures 1 -- 4 respectively, using both the TB (solid and dashed black and red lines) and 14-band \textbf{k}$\cdot$\textbf{p} (solid and dashed green lines) models. The probability densities shown for the TB calculations are averaged over those computed for the ten different supercells (RDs) used to represent each structure. As in our analysis of the bulk electronic structure (cf.~Sec. 3.2), we find that the \textbf{k}$\cdot$\textbf{p} method reliably captures the general trends observed in the full atomistic calculations. However, on average, alloy disorder effects cause a breakdown of the EFA for hole states in GaAs$_{1-x}$Bi$_{x}$ QWs, and for both electron and hole states in GaN$_{y}$As$_{1-x-y}$Bi$_{x}$ QWs. On this basis we conclude that the electronic properties of GaN$_{y}$As$_{1-x-y}$Bi$_{x}$/GaAs QWs are strongly influenced by the alloy microstructure, and hence that the bound states in GaN$_{y}$As$_{1-x-y}$Bi$_{x}$ heterostructures are generally characterised by strong localisation and a corresponding degradation in Bloch character, compared to those in equivalent structures based on conventional semiconductor alloys.


\subsection{Optical transitions in GaN$_{y}$As$_{1-x-y}$Bi$_{x}$/GaAs quantum wells}
\label{sec:results_qw_optical}

We consider now the optical properties of these QW structures and, again facilitated by comparison with the results of the atomistic TB and continuum \textbf{k}$\cdot$\textbf{p} calculations, quantify the impact of alloy disorder on the QW band gap and ground state optical transition strengths.

The results of our TB calculations of the optical properties for structures 1 -- 4 are summarised in Figs.~\ref{fig:optical_transition_strengths} (a) -- 4(d) respectively where, in each case, the results for all ten different RDs are shown. Solid black (red) lines show the computed TE- (TM-) polarised $e1$-$h1$ optical transition strengths, plotted in each case at the wavelength $\lambda$ corresponding to the computed $e1$-$h1$ transition energy. We begin by noting that since all QWs considered are compressively strained, in the \textbf{k}$\cdot$\textbf{p} calculations $h1$ is purely HH-like at the QW zone centre ($\textbf{k}_{\parallel} = 0$). As such, the $h1$ eigenstates computed using the 14-band model have no component along the [001] direction, so that the zone-centre TM-polarised optical transition strength vanishes in accordance with the conventional selection rules for QWs. Examining the results of the atomistic calculations in Figs.~\ref{fig:optical_transition_strengths}(a) and 4(b) we note that this selection rule holds generally in the N-free GaAs$_{1-x}$Bi$_{x}$/GaAs structures 1 and 2: the calculated TM-polarised optical transition strengths are negligibly small, even in the presence of significant alloy disorder at $x = 9$\%. \cite{Usman_APL_2014}


 \begin{figure*}[t!]
 	\includegraphics[width=0.95\textwidth]{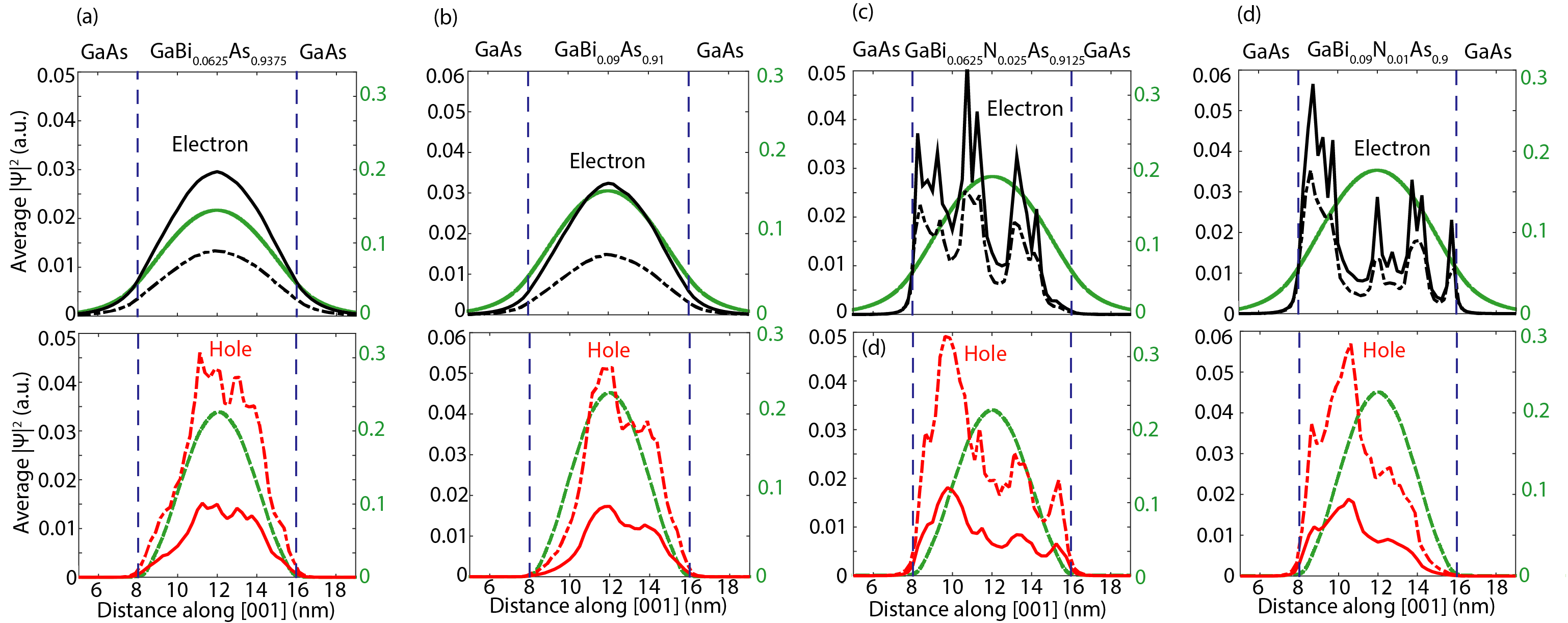}
	\caption{(a), (b), (c) and (d) respectively show the probability density associated with the $e1$ (upper row) and $h1$ (lower row) states in structures 1, 2, 3 and 4 of Table~\ref{tab:qw_structures}. Solid (dash-dotted) black and red lines denote the electron and hole probability densities at cation (anion) sites, calculated using the TB model and averaged over the ten different RDs considered for each structure. Solid (dashed) green lines denote the $e1$ ($hh1$) probability density calculated using the 14-band \textbf{k}$\cdot$\textbf{p} Hamiltonian in the EFA. Dashed blue lines denote the well/barrier interfaces.}
 	\label{fig:qw_probability_density_avg}
 \end{figure*}

The calculated TE-polarised optical transition strengths for structures 1 and 2 describe that the band edge optical transitions in GaAs$_{1-x}$Bi$_{x}$/GaAs QWs are characterised by inhomogeneous broadening associated with alloy disorder-induced fluctuations in the QW band gap, \cite{Usman_PRB_2013,Usman_APL_2014} producing a spectral width $\Delta \lambda$ of the $e1$-$h1$ transition wavelength -- computed as the standard deviation of the $e1$-$h1$ wavelengths for the ten different RDs used to represent each QW -- of 10 nm (17 nm) at $x = 6.25$\% (9\%). \cite{Usman_PRB_2013,Usman_APL_2014} The TB calculations indicate little degradation in the optical transition strengths in going from structure 1 to 2, reflecting that the slight reduction in the Bloch character of $h1$ associated with this $\approx 3$\% increase in $x$ is offset by the increase in electron-hole spatial overlap brought about by the accompanying increase in $\Delta E_{\scalebox{0.6}{\textrm{CB}}}$. \cite{Broderick_IEEEJSTQE_2015} \cite{Usman_PRB_2013,Usman_APL_2014,Broderick_IEEEJSTQE_2015} Indeed, the \textbf{k}$\cdot$\textbf{p} calculations for structures 1 and 2 indicate a modest increase $\lesssim 5$\% in the TE-polarised optical transition strength in going from $x = 6.25$ to 9\%: we attribute the deviation from this trend observed in the TB calculations to the impact of alloy disorder on the $h1$ eigenstates, with the formation of pairs and larger clusters of Bi atoms acting to decrease the Bloch character of the VB edge states. We note that these trends are consistent with the available experimental data, as highlighted in our previous analysis of bulk GaAs$_{1-x}$Bi$_{x}$ alloys and QWs.

Turning our attention to the quaternary QWs -- structures 3 and 4 -- in Figs.~\ref{fig:optical_transition_strengths}(c) and 4(d) we note that co-alloying N and Bi leads to significant modifications of the optical properties. Firstly, we note a breakdown of the conventional QW selection rules: the $h1$ eigenstates in certain supercells acquire appreciable LH character in the presence of N, leading to non-zero zone-centre TM-polarised $e1$-$h1$ optical transition strengths. Our calculations identify this as being primarily a result of the impact of local regions of tensile strain -- due to lattice relaxation about N atomic sites -- on the bound hole states. These microscopic regions of tensile strain allow the predominantly HH-like $h1$ eigenstates to acquire an admixture of LH character -- i.e.~$p$-like orbital components polarised along the growth direction -- and hence to have non-zero transition strengths at the zone centre for TM-polarised transitions involving $e1$. While N incorporation leaves the overall character of the VB edge states largely unchanged on average, this demonstrates that N clustering can nonetheless bring about non-trivial modification of the character of individual bound hole states. This further confirms that calculated breakdown of the conventional optical selection rules is associated with the impact of N-related alloy disorder on the VB structure. This explicitly local effect is not accounted for in the 14-band \textbf{k}$\cdot$\textbf{p} model considered here, nor in existing models of the GaN$_{y}$As$_{1-x-y}$Bi$_{x}$ band structure. We note however that this calculated breakdown of the conventional QW optical selection rules does not necessarily imply that measured optical spectra for GaN$_{y}$As$_{1-x-y}$Bi$_{x}$-based QWs will have large TM-polarised components. Our analysis here focuses on individual QW eigenstates: more detailed analysis shows that this behaviour is associated with the formation of localised states about clusters of N atoms, which occur relatively rarely and are hence unlikely to contribute significantly to the net optical emission/absorption.


\begin{figure}[t!]
 	\includegraphics[width=0.75\columnwidth]{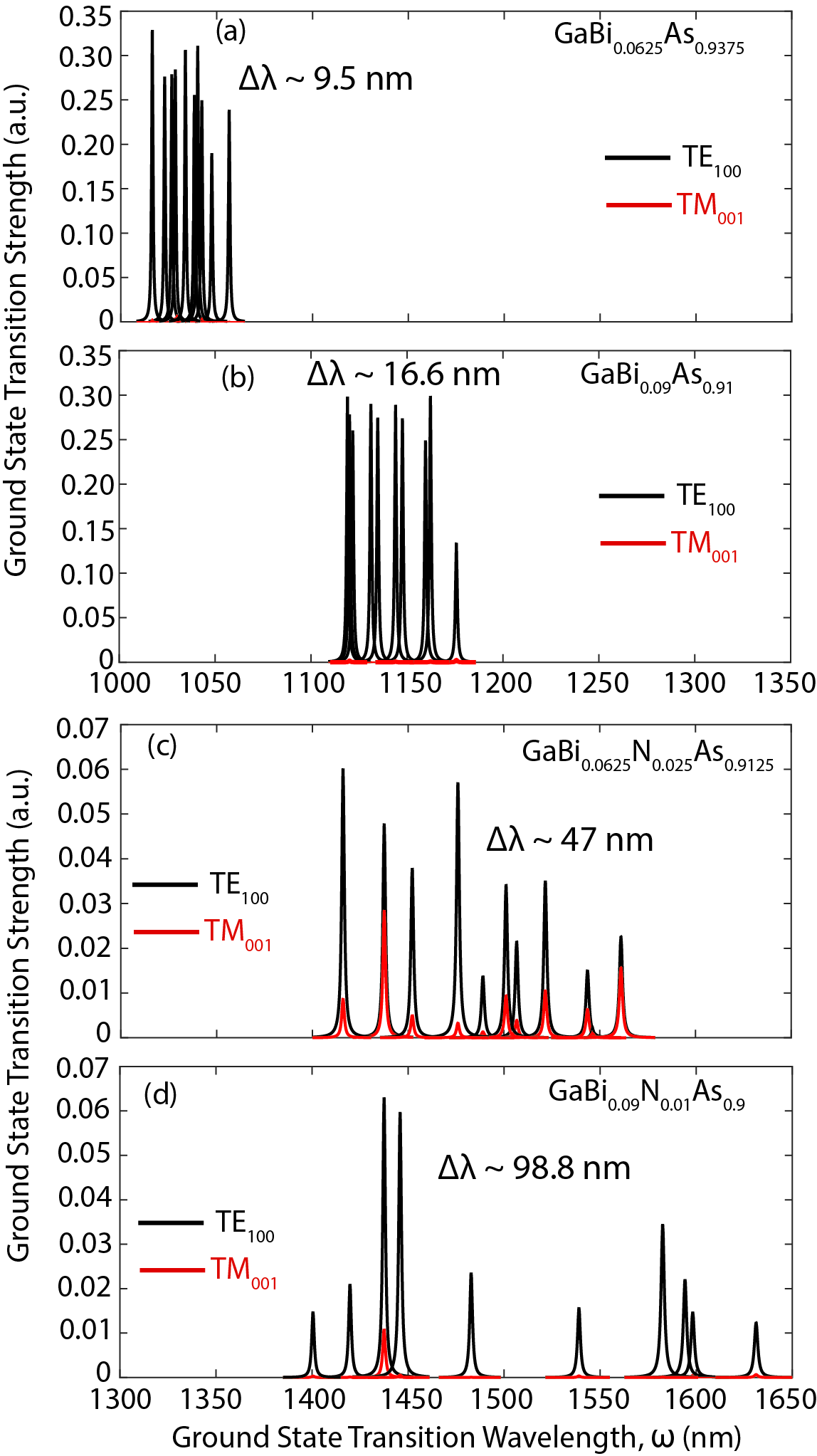}
 	\caption{TE-polarised (solid black lines) and TM-polarised (solid red lines) optical transition strengths between the $e1$ and $h1$ states, calculated using the atomistic TB model, for the N-free GaAs$_{1-x}$Bi$_{x}$/GaAs QWs structure (a) 1, and (b) 2, and for the GaN$_{y}$As$_{1-x-y}$Bi$_{x}$/GaAs QWs structure (c) 3, and (b) 4 (cf.~Table~\ref{tab:qw_structures}). For each structure the computed transition strengths are shown for all ten distinct supercells considered. The spectral width of the ground state transition wavelength ($\Delta \lambda$) is calculated as the standard deviation of the wavelengths computed for the ten distinct supercells (RDs) used to represent each structure.}
 	\label{fig:optical_transition_strengths}
 \end{figure}

Secondly, we note that the combination of N- and Bi-related alloy disorder leads to large variations of the QW band gap between the distinct supercells considered for each structure, resulting in $\Delta \lambda =$ 47 nm (99 nm) in structure 3 (4). For some RDs we calculate anomalously low ground state optical transition strengths, which our analysis associates with strong hybridisation of the $e1$ ($h1$) eigenstates with N (Bi) cluster states that are typically strongly localised, spatially separated from one another, and lie close in energy to the bulk CB or VB edge. We note that the degradation in Bloch character this represents is also manifest through the strong calculated decrease in optical transition strength due to the incorporation of N at fixed $x$ (i.e.~in going from structure 1 to 3, or structure 2 to 4). The \textbf{k}$\cdot$\textbf{p} calculations indicate that the TE-polarised optical transition strength in structure 3 (structure 4) should be $\approx 75$\% ($\approx 80$\%) of that in structure 1 (structure 2), underestimating the approximately fivefold reduction observed in the full atomistic calculations: the extended nature of the QW eigenstates calculated in the EFA does not capture the impact of carrier localisation about N and/or Bi clusters on the $e1$-$h1$ spatial overlap. This discrepancy highlights the role played by short-range alloy disorder in determining the optical properties of GaN$_{y}$As$_{1-x-y}$Bi$_{x}$-based heterostructures, and indicates that intrinsic alloy disorder may potentially limit the optical efficiency and performance of real GaN$_{y}$As$_{1-x-y}$Bi$_{x}$ materials and devices.

The calculated strong variation of the TE-polarised optical transition strengths for the distinct RDs considered reaffirms that the large number of mixed anion local environments that occur in a randomly disordered GaN$_{y}$As$_{1-x-y}$Bi$_{x}$ alloy has a significant impact on the band edge eigenstates. In general, short-range alloy disorder generates a multiplicity of band edge states which are (i) spread over a relatively small range of energies, and (ii) strongly localised, at different locations in the different QW structures investigated. In a real GaN$_{y}$As$_{1-x-y}$Bi$_{x}$ heterostructure the band-edge optical emission/absorption will then consist of contributions from a range of such states, having broadly similar character to the $e1$ and $h1$ states described above. Previous analysis has demonstrated that the net electronic and optical properties arising from such states can be accounted for using EFA-based models by including an appropriate description of the associated inhomogeneous spectral broadening, at least in type-I heterotructures. \cite{Tomic_IEEEJSTQE_2003,Broderick_IEEEJSTQE_2015} This has allowed, e.g., quantitative prediction of the optical gain in (In)GaN$_{y}$As$_{1-y}$ GaAs$_{1-x}$Bi$_{x}$ QW laser structures. \cite{Hader_APL_2000,Marko_SR_2016} As such, while full atomistic calculations demonstrate a breakdown of the EFA for individual eigenstates in GaN$_{y}$As$_{1-x-y}$Bi$_{x}$, and are required in general to understand the full details of the unusual material properties, it is possible that EFA-based models may provide an average description of the electronic and optical properties which accurately describes the measured trends in real heterostructures. However, the stronger localisation of electrons in the N-containing QWs (cf.~Figs.~\ref{fig:qw_probability_density_avg}(c) and \ref{fig:qw_probability_density_avg}(d)) suggest that the EFA may overestimate the carrier spatial overlap in type-II GaN$_{y}$As$_{1-y}$/GaAs$_{1-x}$Bi$_{x}$ QW structures.

Overall, our results indicate that short-range alloy disorder has a marked impact on the GaN$_{y}$As$_{1-x-y}$Bi$_{x}$ alloy properties, similar in nature to equivalent effects in Ga(In)N$_y$As$_{1-y}$ and GaAs$_{1-x}$Bi$_{x}$ alloys and heterostructures.  \cite{Tomic_IEEEJSTQE_2003,Broderick_IEEEJSTQE_2015} We expect that a simliar \textbf{k}$\cdot$\textbf{p}-based approach to that developed for (In)GaN$_{y}$As$_{1-y}$ and (In)GaAs$_{1-x}$Bi$_{x}$ can be reliably applied to compute, analyse and optimise the properties of devices based on type-I GaN$_{y}$As$_{1-x-y}$Bi$_{x}$ heterostructures, but that detailed comparison to future experimental measurements will be required to ascertain the validity of such approaches when applied to type-II heterostructures based on N- and Bi-containing alloys. \cite{Broderick_SR_2017}


\section{Implications for practical applications}
\label{sec:applications}

\begin{figure}[t!]
 	\includegraphics[width=0.80\columnwidth]{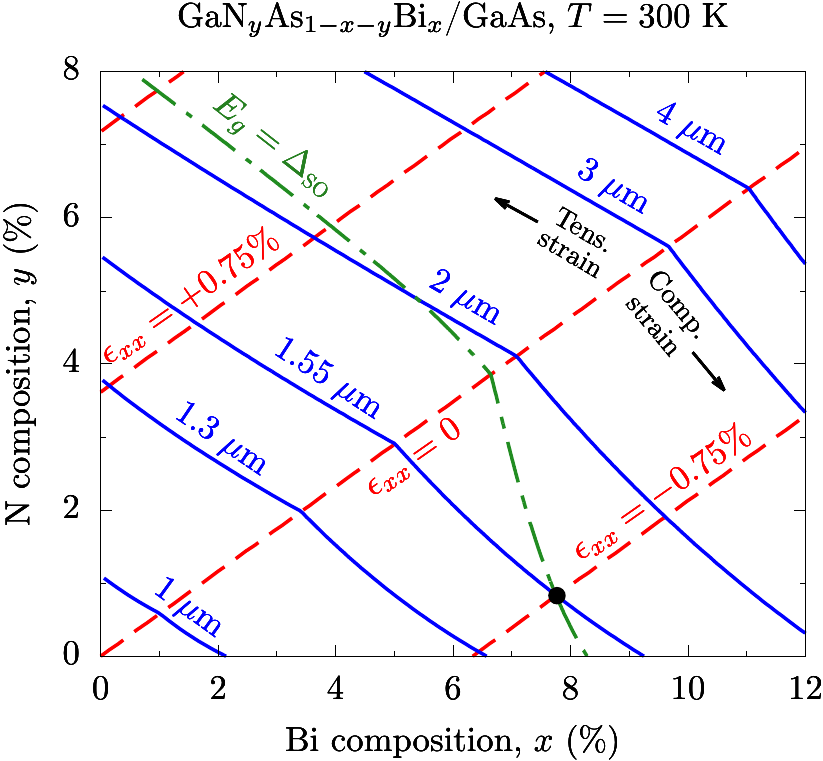}
 	\caption{Composition space map calculated using the 14-band \textbf{k}$\cdot$\textbf{p} Hamiltonian, describing the variation of the in-plane strain ($\epsilon_{xx} = \epsilon_{yy}$) and room temperature band gap ($E_{g}$) in pseudomorphically strained GaN$_{y}$As$_{1-x-y}$Bi$_{x}$ alloys grown on [001]-oriented GaAs. Solid blue and dashed red lines respectively denote paths in the composition space along which $E_{g}$ and $\epsilon_{xx}$ are constant. The dashed-dotted green line denotes alloy compositions for which $E_{g} = \Delta_{\protect\scalebox{0.6}{\textrm{SO}}}$; alloys lying to the right of this contour have $\Delta_{\protect\scalebox{0.6}{\textrm{SO}}} > E_{g}$ and are predicted to have suppressed CHSH Auger recombination and IVBA.}
 	\label{fig:composition_space_map}
\end{figure}

Having quantified key trends in the evolution of the GaN$_{y}$As$_{1-x-y}$Bi$_{x}$ electronic and optical properties, we turn our attention to the consequences of the unusual material band structure for device applications. As we have demonstrated above that the 14-band \textbf{k}$\cdot$\textbf{p} Hamiltonian describes well the composition dependent band edge energies, we employ this model to compute the dependence of the energy gaps on N and Bi composition, and epitaxial strain. These calculations are summarised in Fig.~\ref{fig:composition_space_map}, where solid blue (dashed red) lines denote alloy compositions for which fixed band gap $E_{g}$ (strain $\epsilon_{xx}$) can be achieved in pseudomorphically strained GaN$_{y}$As$_{1-x-y}$Bi$_{x}$/GaAs. The $\epsilon_{xx} = 0$ line describes that lattice-matching to GaAs is achieved for $y = 0.58 \, x$. The extremely strong reduction of $E_{g}$ allows long emission/absorption wavelengths -- ranging from $\sim 1$ $\mu$m through the near-infrared to mid-infrared wavelengths in excess of 4 $\mu$m -- to be achieved at N and Bi compositions compatible with established epitaxial growth. \cite{Yoshimoto_JJAP_2004,Huang_JAP_2005,Tixier_APL_2005,Yoshimoto_JCG_2007,Bushell_JCG_2014,Bushell_JAP_2018}

The dash-dotted green line in Fig.~\ref{fig:composition_space_map} denotes alloy compositions for which $E_{g} = \Delta_{\scalebox{0.6}{\textrm{SO}}}$, so that alloys lying to the left (right) of this contour have $\Delta_{\scalebox{0.6}{\textrm{SO}}} < E_{g}$ ($\Delta_{\scalebox{0.6}{\textrm{SO}}} > E_{g}$). We recall that Auger recombination and IVBA processes involving the SO band play an important role in limiting the performance of near- and mid-infrared light-emitting devices. \cite{Marko_SSMS_2014} For processes involving the SO band there are three distinct cases to consider: (i) $\Delta_{\scalebox{0.6}{\textrm{SO}}} < E_{g}$, (ii) $\Delta_{\scalebox{0.6}{\textrm{SO}}} = E_{g}$, and (iii) $\Delta_{\scalebox{0.6}{\textrm{SO}}} > E_{g}$. Case (i) is typical of near-infrared GaAs- and InP-based light-emitting devices operating at wavelengths between 0.9 and 1.7 $\mu$m. In this regime CHSH Auger recombination and IVBA involving the SO band are present, with the magnitude of these effects increasing strongly as the difference $E_{g} - \Delta_{\scalebox{0.6}{\textrm{SO}}}$ decreases. \cite{Silver_IEEEJQE_1997} In GaAs-based devices operating close to 1 $\mu$m ($E_{g} = 1.24$ eV) one typically has $E_{g} - \Delta_{\scalebox{0.6}{\textrm{SO}}} \approx 0.90$ eV, with CHSH Auger recombination and IVBA playing a minimal role. However, in InP-based devices operating at 1.55 $\mu$m ($E_{g} = 0.8$ eV) is it typical to have $E_{g} - \Delta_{\scalebox{0.6}{\textrm{SO}}} \approx 0.45$ eV, with CHSH Auger recombination and IVBA assuming a dominant role in limiting device performance. \cite{Sweeney_IEEEPTL_1998,Sweeney_PSSB_1999} This trend has been clearly observed in temperature- and pressure-dependent experimental measurements performed on a range of InP-based devices, highlighting that the threshold current density increases superlinearly with emission wavelength as one approaches case (ii). \cite{Sweeney_IEEEISLC_2010,Sweeney_ICTON_2011}

In case (ii), which is denoted by the closed black circle in Fig.~\ref{fig:composition_space_map}, the CHSH Auger recombination and IVBA processes are resonant and can be expected to place severe limitations on efficiency, thereby all but eliminating the potential for sustainable device operation. For case (iii) an electron-hole pair recombining across the band gap provides insufficient energy to promote an electron from the SO band to a VB edge hole state, thereby suppressing CHSH Auger recombination and IVBA by conservation of energy. \cite{Sweeney_ICTON_2011,Broderick_SST_2012} This is the case in GaSb-based mid-infrared devices operating at wavelengths between 2 and 3 $\mu$m where, in line with the expected suppression of CHSH Auger recombination and IVBA, reduced threshold current densities compared to those in equivalent InP-based near-infrared devices are observed in experiment. \cite{Sweeney_ICTON_2011}

As such, there are two desirable scenarios from the perspective of the relationship between $E_{g}$ and $\Delta_{\scalebox{0.6}{\textrm{SO}}}$. Firstly, when $\Delta_{\scalebox{0.6}{\textrm{SO}}} < E_{g}$ with the ratio $\frac{ \Delta_{\scalebox{0.6}{\textrm{SO}}} }{ E_{g} }$ significantly less than one, in which case CHSH Auger recombination and IVBA occur but at sufficiently low rates so as not to impede device performance. Secondly, the ideal scenario in which $\Delta_{\scalebox{0.6}{\textrm{SO}}} > E_{g}$ and CHSH Auger recombination and IVBA can be expected to be suppressed. While the ideal $\Delta_{\scalebox{0.6}{\textrm{SO}}} > E_{g}$ scenario can in principle be achieved in GaAs$_{1-x}$Bi$_{x}$ ($y = 0$) with $E_{g} = 1.55$ $\mu$m, our calculations show that incorporating N requires a reduction in $x$ to maintain a fixed emission wavelength, and therefore reduces $\Delta_{\scalebox{0.6}{\textrm{SO}}}$ relative to $E_{g}$: an undesirable change which pushes the band structure from case (iii) towards case (ii). N incorporation then quickly brings $E_{g}$ and $\Delta_{\scalebox{0.6}{\textrm{SO}}}$ into resonance -- i.e.~case (ii), $E_{g} = \Delta_{\scalebox{0.6}{\textrm{SO}}}$ -- the worst possible scenario in terms of the suitability of the band structure for the development of a light-emitting device.


The results in Fig.~\ref{fig:composition_space_map} then indicate that N-free GaAs$_{1-x}$Bi$_{x}$ alloys provide the ideal scenario for the realisation of efficient light emission at 1.55 $\mu$m, and that co-alloying N with Bi to achieve 1.55 $\mu$m emission presents a band structure which is not compatible with the design of light-emitting devices. As a result of the strong increase in $\Delta_{\scalebox{0.6}{\textrm{SO}}}$ brought about by Bi incorporation, at 1.3 $\mu$m alloys containing Bi will have $\Delta_{\scalebox{0.6}{\textrm{SO}}} < E_{g}$ but with the difference $E_{g} - \Delta_{\scalebox{0.6}{\textrm{SO}}}$ being smaller than in a conventional InP-based material. This suggests that increased losses due to CHSH Auger recombination and IVBA can be expected in a 1.3 $\mu$m GaAs-based GaN$_{y}$As$_{1-x-y}$Bi$_{x}$ device compared to a conventional InP-based device operating at the same wavelength.

Additionally, we note that it is generally desirable to exploit strain in QW lasers to enhance performance via reduction of the density of states close in energy to the VB edge (in compressively strained structures), or via polarisation selectivity (in tensile strained structures). \cite{Reilly_IEEEJQE_1994,Adams_IEEEJSTQE_2011} At fixed emission wavelength one usually seeks to utilise an alloy having as large a strain $\vert \epsilon_{xx} \vert$ -- compressive or tensile -- as is compatible with epitaxial growth. The decrease in compressive strain compared to that in GaAs$_{1-x}$Bi$_{x}$ associated with incorporating N {while maintaining a fixed emission wavelength can be expected to reduce the performance of a QW laser structure compared to that expected for an equivalent GaAs$_{1-x}$Bi$_{x}$/GaAs laser structure. \cite{Broderick_IEEEJSTQE_2015} Therefore, the expected degradation in the suitability of the band structure in utilising quaternary GaN$_{y}$As$_{1-x-y}$Bi$_{x}$ in favour of (i) a conventional InP-based (Al)In$_{1-x}$Ga$_{x}$As heterostructure to achieve 1.3 $\mu$m emission, or (ii) a ternary GaAs$_{1-x}$Bi$_{x}$ heterostructure to achieve 1.55 $\mu$m emission suggests, contrary to recent analysis, \cite{Habchi_MSSP_2014,Fan_JAP_2016} that GaN$_{y}$As$_{1-x-y}$Bi$_{x}$ alloys and heterostructures containing N are \textit{not} suitable for applications at 1.3 or 1.55 $\mu$m.

Our analysis does however confirm the potential of GaN$_{y}$As$_{1-x-y}$Bi$_{x}$ alloys for applications in multi-junction solar cells, \cite{Sweeney_IEEEPVSC_2013} since $E_{g} \approx 1$ eV can be achieved in alloys which can be grown lattice-matched to either GaAs or Ge ($y = 0.58 \, x$, $x \approx 2.9$\%). While the difference $E_{g} - \Delta_{\scalebox{0.6}{\textrm{SO}}}$ for such alloys would not be as favourable as in an InP-based material designed to have a similar absorption wavelength, it is expected to be sufficiently large so as not to impede performance. Furthermore, the carrier densities present in an illuminated solar cell are significantly lower than those in an electrically pumped semiconductor laser, making Auger recombination -- which scales roughly as the cube of the injected carrier density -- less of an impediment to device operation. The modest N and Bi compositions required to reach a band gap $\sim 1$ eV have been achieved in initial growth studies of GaN$_{y}$As$_{1-x-y}$Bi$_{x}$/GaAs epitaxial layers. \cite{Yoshimoto_JJAP_2004,Huang_JAP_2005,Tixier_APL_2005,Yoshimoto_JCG_2007,Bushell_JCG_2014,Bushell_JAP_2018} Analysis of prototypical GaAs$_{1-x}$Bi$_{x}$/GaAs QW solar cells indicates that one potential factor limiting the photovoltaic performance of lattice-matched GaN$_{y}$As$_{1-x-y}$Bi$_{x}$ junctions may be the large inhomogeneous spectral broadening of the band edge optical absorption, associated with the presence of short-range alloy disorder and crystalline defects. \cite{Wilson_EPSECE_2017} However, we expect that these issues could be mitigated to some degree via a combination of refinement of the epitaxial growth, \cite{Luo_PRB_2015,Luo_NPGAM_2017} sample preparation, \cite{Mazzucato_SST_2013,Makhloufi_NRL_2014} and device design. This approach has been successfully employed to develop multi-junction solar cells incorporating the dilute nitride alloy (In)GaN$_{y}$As$_{1-y}$, which have demonstrated record-breaking efficiency. \cite{Sabnit_AIPCP_2012}

Finally, since our calculations indicate that GaN$_{y}$As$_{1-x-y}$Bi$_{x}$ alloys display effectively identical enhancement of the spin-orbit coupling to that in GaAs$_{1-x}$Bi$_{x}$, similarly large enhancement of the Rashba spin-orbit interaction \cite{Simmons_APL_2015} to that in GaAs$_{1-x}$Bi$_{x}$ can be expected in GaN$_{y}$As$_{1-x-y}$Bi$_{x}$, potentially opening up applications in spintronic devices. \cite{Mazzucato_APL_2013}


\section{Conclusions}
\label{sec:conclusions}

We have developed a multi-scale theoretical framework to calculate the properties of GaN$_{y}$As$_{1-x-y}$Bi$_{x}$ highly-mismatched alloys and heterostructures, based on carefully derived atomistic TB and continuum \textbf{k}$\cdot$\textbf{p} Hamiltonians. We have performed a systematic investigation revealing key trends in the electronic and optical properties of bulk GaN$_{y}$As$_{1-x-y}$Bi$_{x}$ alloys and GaN$_{y}$As$_{1-x-y}$Bi$_{x}$/GaAs QWs. Our analysis indicates that GaN$_{y}$As$_{1-x-y}$Bi$_{x}$ alloys provide broad scope for band structure engineering: incorporating N (Bi) allows to manipulate the CB (VB) structure close in energy to band edges, offering significant control over the band gap, VB spin-orbit splitting energy, and band offsets. Since incorporating N (Bi) brings about tensile (compressive) strain with respect to a GaAs substrate, co-alloying N and Bi then further delivers significant control over the strain in epitaxial layers and heterostructures, providing further opportunities to tailor the electronic and optical properties. The intrinsic flexibility of the GaN$_{y}$As$_{1-x-y}$Bi$_{x}$ alloy band structure is therefore particularly appealing for practical applications.

Through systematic analysis of large-scale atomistic electronic structure calculations we demonstrated that the respective impact of N and Bi incorporation on the CB and VB structure remain effectively independent, even in the presence of significant short-range alloy disorder. Comparison of atomistic and continuum calculations highlights that a 14-band \textbf{k}$\cdot$\textbf{p} Hamiltonian is sufficient to describe the evolution of the main features of the GaN$_{y}$As$_{1-x-y}$Bi$_{x}$ band structure, with the predicted evolution of the band gap and VB spin-orbit splitting energy in good agreement with experimental measurements. Applying the TB model to compute the electronic and optical properties of realistically sized GaN$_{y}$As$_{1-x-y}$Bi$_{x}$/GaAs QWs demonstrates that short-range alloy disorder produces strong carrier localisation, ultimately leading to significant inhomogeneous spectral broadening as well as a breakdown of the conventional selection rules governing QW optical transitions. On this basis we conclude that alloy disorder effects are likely to play an important role in determining the properties of real GaN$_{y}$As$_{1-x-y}$Bi$_{x}$ alloys and heterostructures. While our analysis suggests that \textbf{k}$\cdot$\textbf{p}-based models incorporating appropriately parametrised inhomogeneous spectral broadening are likely to be suitable when applied to type-I heterostructures, further detailed analysis of N- and Bi-containing type-II heterostructures -- e.g.~GaN$_{y}$As$_{1-y}$/GaAs$_{1-x}$Bi$_{x}$ QWs -- may require a theoretical description based upon direct atomistic calculations.

Based on our analysis of the electronic structure we discussed implications of the unusual properties of GaN$_{y}$As$_{1-x-y}$Bi$_{x}$ alloys for practical applications. Contrary to the existing literature our analysis suggests that GaN$_{y}$As$_{1-x-y}$Bi$_{x}$ heterostructures are not suitable for the development of GaAs-based semiconductor lasers operating at 1.3 or 1.55 $\mu$m: the reduction in compressive strain and spin-orbit splitting energy compared to equivalent N-free GaAs$_{1-x}$Bi$_{x}$ structures is expected to compromise the proposed benefits of Bi incorporation and hence place significant limitations on performance. Overall, we conclude that the features of the electronic structure revealed through our analysis indicate that GaN$_{y}$As$_{1-x-y}$Bi$_{x}$ alloys are most promising as a suitable $\sim$ 1 eV band gap material for applications in multi-junction solar cells, where they can provide GaAs- and Ge-compatible lattice- and current-matched junctions.


\section*{Acknowledgements}

M.U.~and C.A.B.~contributed equally to this work. This work was supported by the European Commission (project no.~FP7-257974), by Science Foundation Ireland (SFI; project no.~15/IA/3082), and by the Engineering and Physical Sciences Research Council, U.K. (EPSRC; project no.~EP/K029665/1). M.U.~acknowledges the use of computational resources from the National Science Foundation (NSF, U.S.A.) funded Network for Computational Nanotechnology (NCN) through \url{http://nanohub.org}. The authors thank Dr.~Zoe L.~Bushell and Prof.~Stephen J.~Sweeney of the University of Surrey, U.K., for useful discussions and for providing access to the results of their experimental measurements prior to publication.

\newpage
\clearpage 

\renewcommand{\thefigure}{S\arabic{figure}}
\setcounter{figure}{0}

\renewcommand{\thetable}{S\arabic{table}}
\setcounter{table}{0}

\renewcommand{\theequation}{S\arabic{equation}}
\setcounter{equation}{0}

\begin{widetext}

\Large \textbf{\textit{\underline{Supplementary Material Section}}}
\\ \\
\end{widetext}

\noindent
\\ \\

\normalsize


\vspace{-1.0cm}
\noindent
\textbf{Section S1: Theoretical Models}
\\ \\
\noindent
The unusual electronic properties of dilute nitride and bismide alloys derive from the fact that, when incorporated in dilute concentrations, N and Bi act as isovalent impurities which strongly perturb the band structure of the host matrix semiconductor. \cite{Wei_PRL_1996,Kent_PRB_2001,Lindsay_SSC_2001} As a result, the details of electronic structure are strongly influenced, even at dilute compositions, by short-range alloy disorder \cite{Kent_PRB_2001,Reilly_JPCM_2004,Lindsay_PRL_2004,Usman_PRB_2013} -- i.e.~the formation of pairs and larger clusters of N and Bi atoms sharing common cation nearest neighbours -- meaning that highly-mismatched alloys pose significant challenges for theory. Firstly, due to the prominence of impurity effects conventional approaches to analyse alloy band structures, such as the virtual crystal approximation, break down. Secondly, since the effects of Bi and N incorporation are prominent at dilute compositions and in the presence of alloy disorder, quantitative understanding of the properties of real materials must be built on analysis of systems containing upwards of thousands of atoms: \cite{Kent_PRB_2001,Reilly_SST_2002,Zhang_PRB_2005,Usman_PRB_2011} the minimum N or Bi composition that can be considered using a 2$M$-atom Ga$_{M}$As$_{M-1}$X$_{1}$ (X = N, Bi) supercell is $M^{-1}$. For example, to consider an alloy composition of $0.1$\% then requires a supercell containing a minimum of $2000$ atoms ($M^{-1} = 10^{-3}$) and, since a 2000-atom supercell corresponds only to a single substitutional impurity, there is no scope to investigate disorder effects unless the system size is significantly increased.

These features generally place the study of highly-mismatched alloys beyond the reach of contemporary first principles methods, since sufficiently advanced approaches cannot be applied \textit{directly} to such large systems due to their associated computational expense. This then mandates the development of alternative theoretical approaches. Indeed, while first principles analyses have provided valuable insights and information, detailed understanding of the properties of real dilute nitride and bismide alloys has to date primarily been built via the development and application of appropriate semi-empirical pseudopotential \cite{Kent_PRB_2001,Kent_PRL_2001,Kent_SST_2002,Zhang_PRB_2005} and tight-binding (TB) \cite{Reilly_SST_2002,Reilly_JPCM_2004,Lindsay_PRL_2004,Shtinkov_PRB_2003,Usman_PRB_2011,Broderick_PRB_2014} models. Further analysis has shown that simple continuum models -- derived and parametrised on the basis of either (i) experimental measurements, \cite{Shan_PRL_1999,Alberi_PRB_2007} or (ii) electronic structure calculations \cite{Reilly_SST_2002,Broderick_SST_2013} -- provide an effective means to understand and analyse the main features of the band structure of N- and Bi-containing alloys, by describing general trends in important material parameters (despite omitting the detailed features of the electronic structure). Here, in the following two sections, we present two models of the electronic structure of GaN$_{y}$As$_{1-x-y}$Bi$_{x}$ alloys: (i) an atomistic $sp^{3}s^{*}$ TB model, and (ii) a continuum, extended basis set 14-band \textbf{k}$\cdot$\textbf{p} Hamiltonian.
\\ \\
\textbf{Section S1.A: Atomistic Tight-binding Model}
\\ \\
\noindent
Since the TB method employs a basis set of localised atomic orbitals, it is ideally suited to probe the electronic structure of localised impurities. \cite{Reilly_JPCS_2010} This, combined with its low computational cost, means that appropriately parametrised TB models provide a physically transparent and highly effective means by which to systematically analyse the properties of large, disordered systems. We have previously demonstrated that the TB method provides a detailed understanding of the electronic and optical properties of GaN$_{y}$As$_{1-y}$ and GaAs$_{1-x}$Bi$_{x}$ alloys, and that the results of calculations based on this approach are in quantitative agreement with a wide range of experimental data. \cite{Lindsay_PRL_2004,Reilly_SST_2009,Usman_PRB_2011,Usman_PRB_2013,Broderick_PRB_2014} Here, we extend this approach to quaternary GaN$_{y}$As$_{1-x-y}$Bi$_{x}$ alloys.

To accurately describe states lying close in energy to the VB and CB edges we employ an $sp^{3}s^{*}$ basis set, \cite{Vogl_JPCS_1983} thereby allowing the dispersion of the lowest CB to be described throughout the entire Brillouin zone without explicit introduction of $d$ orbitals, which avoids significant parametric complexity. Our TB parameters for the band structures of the GaN, GaAs and GaBi compounds are obtained on the basis of first principles band structure calculations. \cite{Janotti_PRB_2002,Usman_PRB_2011} Spin-orbit coupling between $p$ orbitals is included explicitly in the model, in order to account for the known, strong relativistic effects associated with Bi incorporation in (In)GaAs. \cite{Fluegel_PRL_2006,Usman_PRB_2011,Batool_JAP_2012} To construct alloy supercells we replace As atoms by N or Bi atoms at selected positions on the anion sublattice. To account for the important effects \cite{Kent_PRB_2001,Lindsay_PB_2003,Zhang_PRB_2005,Usman_PRB_2011} of lattice relaxation about N and Bi lattice sites, the relaxed atomic positions in a given supercell are determined by minimising the total elastic energy using a valence force field (VFF) model, based on the Keating potential and including anharmonic corrections to account accurately for the large local strains arising due to the significant differences in the covalent radii of N, As and Bi. \cite{Keating_PR_1966,Lazarenkova_APL_2004,Usman_PRB_2011}

In our TB model the atomic orbital energies for a given atom are taken to depend on the local neighbour environment by (i) averaging over the orbital energies of the compounds formed by the atom and each of its nearest neighbours, and (ii) renormalising the nearest-neighbour averaged orbital energy obtained in this manner to account firstly for the mixed-anion local environment (seen by a cation having nearest neighbour anions of different atomic species) and, secondly, for the non-local nature of the lattice relaxation brought about by the aforementioned large differences in the anion covalent radii. In this approach the energy of orbital $\alpha$ ($= s, p_{x}, p_{y}, p_{z}, s^{*}$) in the atom located at lattice site $n$ ($= 1, \dots, 2M$) is given by \cite{Lindsay_thesis_2002}

\begin{widetext}

\begin{eqnarray}
	E_{n\alpha} &=& \frac{1}{4} \, \sum_{j} \bigg[ \underbrace{ \Delta E_{\scalebox{0.6}{\textrm{VB}}} ( nj ) + E_{\alpha} ( nj ) }_{ \textrm{Nearest-neighbour bond} } + \underbrace{ \sum_{k \neq j} \left( \frac{ \Delta E_{\alpha} }{ \Delta d_{0} } \right)_{njk} \bigg( d ( jk ) - d_{0} ( jk ) \bigg) }_{ \textrm{Mixed anion local environment} } \nonumber \\
	&+& \underbrace{ \left( \frac{ \sum_{\textrm{X}} M_{\scalebox{0.6}{\textrm{X}}} \left( \frac{ \Delta E_{\alpha} }{ \Delta d_{0} } \right)_{\textrm{GaAsX}} }{ \sum_{\textrm{X}} M_{\scalebox{0.6}{\textrm{X}}}  } \right) \bigg( d ( nj ) - d_{0} ( nj ) \bigg) }_{ \textrm{Non-local lattice relaxation} } \bigg] \, , \label{eq:orbital_energy}
\end{eqnarray}

\end{widetext}

\noindent
where X = N or Bi, and the sum runs over the four nearest neighbours $j$ of atom $n$. $E_{\alpha} ( nj )$, $\Delta E_{\scalebox{0.6}{\textrm{VB}}} ( nj )$ and $d_{0} ( nj )$ respectively denote the energy of orbital $\alpha$, the VB offset, and the unstrained nearest-neighbour bond length of the compound formed by atoms $n$ and $j$. $d ( nj )$ is the relaxed bond length between atoms $n$ and $j$ in the supercell, $M_{\scalebox{0.6}{\textrm{X}}}$ is the total number of N \textit{or} Bi atoms, and $\sum_{\textrm{X}} M_{\scalebox{0.6}{\textrm{X}}}$ is the total number of N \textit{and} Bi atoms. The magnitude of the second and third terms in Eq.~\eqref{eq:orbital_energy} is determined by \cite{Lindsay_thesis_2002}

\begin{equation}
	\bigg( \frac{ \Delta E_{\alpha} }{ \Delta d_{0} } \bigg)_{njk} = \frac{ E_{\alpha} ( nj ) - E_{\alpha} ( nk ) }{ d_{0} ( nj ) - d_{0} ( nk ) } \, ,
	\label{eq:difference_ratio}
\end{equation}

\noindent
which is the ratio of the difference between the energy of orbital $\alpha$ to that between the equilibrium bond lengths, for the compounds formed by atom $n$ and its nearest neighbours $j$ and $k$ (with $E_{\alpha}$ assumed here to include the corresponding VB offset $\Delta E_{\scalebox{0.6}{\textrm{VB}}}$ for a given compound).

The first term in Eq.~\eqref{eq:orbital_energy} describes conventional averaging of the orbital energy over nearest-neighbours. The second and third terms in Eq.~\eqref{eq:orbital_energy} describe a renormalisation of the orbital energy, and respectively account for (i) the large, bond length dependent differences between the orbital energies in the compounds formed by a given atom and nearest neighbours of differing atomic species, and (ii) the relaxation of the crystal lattice in response to substitutional incorporation of an impurity X atom. Previous analysis \cite{Lindsay_thesis_2002,Lindsay_SSE_2003} has shown that it is pertinent to include an orbital energy renormalisation of this form to account for the fact that substitutional N and Bi atoms have significant differences in electronegativity and covalent radii compared to the As atoms they replace, meaning that they perturb the structural and electronic properties much more severely than in conventional III-V alloys.

\begin{table}[t!]
 	\caption{\label{tab:tb_parameters_1} Band structure parameters for the constituent compounds in the $sp^{3}s^{*}$ tight-binding model of GaN$_{y}$As$_{1-x-y}$Bi$_{x}$, including the valence band offsets $\Delta E_{\protect\scalebox{0.6}{\textrm{VB}}}$, orbital energies $E_{\alpha}$ ($\alpha = s, p$ and $s^{*}$), inter-atomic interaction matrix elements $V_{\delta}$ ($\delta = ss\sigma$, $pp\sigma$, $pp\pi$, $s_{c}p_{a}\sigma$, $s_{a}p_{c}\sigma$, $s_{c}^{*}p_{a}\sigma$ and $s_{a}^{*}p_{c}\sigma$), and renormalised atomic spin-orbit splitting energies $\Delta_{c,a}$. All parameters are given in units of eV.}
 	\begin{ruledtabular}
 		\begin{tabular}{cccc}
 		Parameter & GaN & GaAs & GaBi \\
 			\hline

 			$\Delta E_{\scalebox{0.6}{\textrm{VB}}}$ & $-2.28$ &  $0.00$ &  $1.10$ \\

 			\hline

 			$E_{s_{c}    }$ & $-0.9994$  & $-2.9474$ & $-5.6126$ \\
 			$E_{s_{a}    }$ & $-12.3306$ & $-8.6336$ & $-8.3774$ \\
 			$E_{p_{c}    }$ & $~8.5803$  & $~3.5532$ & $~1.6940$ \\
 			$E_{p_{a}    }$ & $-2.7197$  & $~0.9252$ & $-0.1256$ \\
 			$E_{s_{c}^{*}}$ & $12.2000$  & $~6.2000$ & $~5.8164$ \\
 			$E_{s_{a}^{*}}$ & $12.2000$  & $~7.0914$ & $~6.1262$ \\

 			\hline

 			$V_{s        s    \sigma}$ & $-2.0700$ & $-1.6835$ & $-1.3425$ \\
 			$V_{p        p    \sigma}$ & $~5.0530$ & $~2.9500$ & $~2.0003$ \\
 			$V_{p        p    \pi   }$ & $-0.7150$ & $-0.7420$ & $-0.6345$ \\
 			$V_{s_{c}    p_{a}\sigma}$ & $~3.0593$ & $~2.4200$ & $~1.2025$ \\
 			$V_{s_{a}    p_{c}\sigma}$ & $~2.0717$ & $~2.3920$ & $~2.3567$ \\
 			$V_{s_{c}^{*}p_{a}\sigma}$ & $~2.8375$ & $~2.0400$ & $~0.5100$ \\
 			$V_{s_{a}^{*}p_{c}\sigma}$ & $~1.5442$ & $~1.7700$ & $~1.8051$ \\

 			\hline

 			$\Delta_{c}$ & $0.0020$ & $0.1350$ & $0.6714$ \\
 			$\Delta_{a}$ & $0.0070$ & $0.0550$ & $0.0384$ \\
 		\end{tabular}
 	\end{ruledtabular}
 \end{table}


\begin{table}[t!]
 	\caption{\label{tab:tb_parameters_2} Dimensionless exponents $\eta_{\delta}$ ($\delta = ss\sigma$, $pp\sigma$, $pp\pi$, $s_{c}p_{a}\sigma$, $s_{a}p_{c}\sigma$, $s_{c}^{*}p_{a}\sigma$ and $s_{a}^{*}p_{c}\sigma$) used to rescale the inter-atomic interaction matrix elements $V_{\delta}$ for the constituent compounds in the $sp^{3}s^{*}$ tight-binding model of GaN$_{y}$As$_{1-x-y}$Bi$_{x}$ (cf.~Eq.~\eqref{eq:matrix_element_rescaling}).}
 	\begin{ruledtabular}
 		\begin{tabular}{cccc}
 			Parameter & GaN & GaAs & GaBi \\
 			\hline
 			$\eta_{s        s    \sigma}$ & $3.030$ & $3.512$ & $3.660$ \\
 			$\eta_{p        p    \sigma}$ & $2.090$ & $3.204$ & $2.200$ \\
 			$\eta_{p        p    \pi   }$ & $3.720$ & $4.326$ & $3.240$ \\
 			$\eta_{s_{c}    p_{a}\sigma}$ & $3.500$ & $4.500$ & $4.090$ \\
 			$\eta_{s_{a}    p_{c}\sigma}$ & $4.000$ & $4.100$ & $4.080$ \\
 			$\eta_{s_{c}^{*}p_{a}\sigma}$ & $6.200$ & $7.200$ & $7.200$ \\
 			$\eta_{s_{a}^{*}p_{c}\sigma}$ & $4.000$ & $4.200$ & $4.200$ \\
 		\end{tabular}
 	\end{ruledtabular}
 \end{table}


 \begin{table}[t!]
 	\caption{\label{tab:vff_parameters} Unstrained bond lengths ($d_{0}$), and bond stretching ($\alpha$) and bond-angle bending ($\beta$) force constants for the constituent compounds in the valence force field potential used to relax the atomic positions in GaN$_{y}$As$_{1-x-y}$Bi$_{x}$ supercells.}
 	\begin{ruledtabular}
 		\begin{tabular}{cccc}
 			Parameter & GaN & GaAs & GaBi \\
 			\hline
 			$d_{0}$  (\AA)        & $1.944$  & $2.448$ & $2.740$ \\
 			$\alpha$ (N m$^{-1}$) & $158.00$ & $41.18$ & $61.00$ \\
 			$\beta$  (N m$^{-1}$) & $14.66$  & $~8.95$ & $~6.21$ \\
 		\end{tabular}
 	\end{ruledtabular}
 \end{table}

The nature of the orbital energy renormalisation is best understood by considering its action in the presence of an isolated substitutional impurity. In a Ga$_{M}$As$_{M-1}$X$_{1}$ supercell a cation (Ga atom) having the X atom as a nearest neighbour sits in a mixed-anion nearest-neighbour environment (formed by the X atom and three As atoms). Due to relaxation of the crystal lattice, the length of the Ga-X bond is smaller (larger) than the three Ga-As nearest-neighbour bonds when X = N (Bi). \cite{Kent_PRB_2001,Usman_PRB_2011} The second term in Eq.~\eqref{eq:orbital_energy} proceeds in a pairwise manner, adjusting the contribution of a given nearest-neighbour bond to the Ga orbital energy by accounting for the large differences in Ga orbital energies between GaAs and GaX -- i.e.~between those in compounds formed by the Ga atom and its X or As nearest neighbours. For a given nearest-neighbour bond the magnitude of the renormalisation term is determined by the multiplicative factor $d ( jk ) - d_{0} ( jk )$, which is non-zero only when the relaxed bond length differs from $d_{0} ( jk )$, thus describing a simple linear variation of the Ga orbital energy with bond length in going from GaAs to GaX.

The relaxation of the crystal lattice about the X atomic site generally propagates through the crystal on a length scale which exceeds typical nearest-neighbour bond lengths. However, by definition the action of the second term in Eq.~\eqref{eq:orbital_energy} is explicitly limited to nearest neighbours: the As second-nearest neighbours of the X atom in a Ga$_{M}$As$_{M-1}$X$_{1}$ supercell ($M_{\scalebox{0.6}{\textrm{X}}} = \sum_{\scalebox{0.6}{\textrm{X}}} M_{\scalebox{0.6}{\textrm{X}}} = 1$) have four Ga nearest neighbours, in which case the factor defined by Eq.~\eqref{eq:difference_ratio} vanishes, despite that the relaxed second-nearest neighbour bond lengths will be different from those in unstrained GaAs. This non-local perturbation of the crystal structure is taken into account by the third term in Eq.~\eqref{eq:orbital_energy}, which renormalises the orbital energy via a weighted average of the ratio defined in Eq.~\eqref{eq:difference_ratio} over all substitutional impurities in the supercell. The magnitude of this correction is again determined by the difference between the relaxed and unstrained nearest-neighbour bond lengths, meaning that (i) it applies at all atomic sites that have been affected by the lattice relaxation, and (ii) its magnitude decreases as the distance from the X atomic site increases, in line with the magnitude of the lattice relaxation.

This orbital energy renormalisation therefore provides a suitable manner by which to explicitly incorporate the non-local nature of the perturbation to the supercell Hamiltonian -- associated with N and/or Bi incorporation -- into what is implicitly a nearest-neighbour model. Also, since this approach relies only on the known atomic orbital energies and nearest-neighbour bond lengths, it has the benefit of circumventing the need to introduce additional parameters to describe these important beyond-nearest-neighbour effects.

The inter-atomic interaction matrix elements between a given pair of nearest neighbour atoms are computed in the model by (i) using the two-centre integrals of Slater and Koster \cite{Slater_PR_1954} to take into account changes in bond angle, and (ii) incorporating a bond length dependent rescaling to account for the difference between the relaxed bond length between a pair of neighbouring atoms in the supercell, and the equilibrium bond length of the compound formed by the same two atoms. For each distinct type of interaction, the bond length dependent rescaling of the interaction matrix element between nearest neighbours $n$ and $j$ is given by

\begin{equation}
	V_{\delta} ( d ( nj ) ) = V_{\delta} ( nj ) \, \left( \frac{ d_{0} ( nj ) }{ d ( nj ) } \right)^{ \eta_{\delta} } \, ,
	\label{eq:matrix_element_rescaling}
\end{equation}

\noindent
where $V_{\delta} ( nj )$ is the corresponding interaction matrix element in the unstrained compound formed by atoms $n$ and $j$, and $\delta = ss\sigma$, $pp\sigma$, $pp\pi$, $s_{c}p_{a}\sigma$, $s_{a}p_{c}\sigma$, $s_{c}^{*}p_{a}\sigma$ or $s_{a}^{*}p_{c}\sigma$ describes the symmetry of the interaction (with $c$ and $a$ denoting cations and anions, respectively). The exponent $\eta_{\delta}$ takes a distinct value for each $\delta$ -- the full set are determined for each compound by fitting to the hydrostatic deformation potentials obtained from first principles calculations. \cite{Reilly_SST_2009}

In the TB method we compute the momentum matrix element between the supercell eigenstates $\vert n \textbf{k} \sigma \rangle$ and $\vert m \textbf{k} \sigma' \rangle$, at wave vector $\textbf{k}$ and having spins $\sigma$ and $\sigma'$, in the position basis as

\begin{widetext}

\begin{eqnarray}
	P_{nm}^{(\widehat{e})} ( \textbf{k} ) &=& - \frac{ i m_{0} }{ \hbar } \sum_{ \sigma, \sigma' } \langle n \textbf{k} \sigma \vert \left[ \widehat{ \left( \widehat{e} \cdot \textbf{r} \right) }, \widehat{H} \right] \vert m \textbf{k} \sigma' \rangle \nonumber \\
	&=& - \frac{ i m_{0} }{ \hbar } \sum_{ \sigma, \sigma' } \sum_{ p, q } \sum_{ \alpha, \beta } a_{n p \alpha \sigma }^{*} ( \textbf{k} ) a_{m q \beta \sigma'} ( \textbf{k} ) \left( \widehat{e} \cdot \textbf{r}_{p} - \widehat{e} \cdot \textbf{r}_{q} \right) \, \langle p \alpha \sigma \vert \widehat{H} \vert q \beta \sigma' \rangle \, , \label{eq:momentum_tb}
\end{eqnarray}

\end{widetext}

\noindent
where $p$ and $q$ index the atomic sites in the supercell, $\alpha$ and $\beta$ index the atomic orbitals localised at each atomic site, and $\widehat{e}$ is a unit vector defining the direction along which the corresponding emitted/absorbed photon is polarised. The final expression for $P_{nm}^{(\widehat{e})} ( \textbf{k} )$ is obtained using the spectral resolution of the position operator, $\widehat{ \left( \widehat{e} \cdot \textbf{r} \right) } = \sum_{p, \alpha, \sigma} ( \widehat{e} \cdot \textbf{r}_{p} ) \, \vert p \alpha \sigma \rangle \langle p \alpha \sigma \vert$, in addition to assuming that there is no overlap between basis states localised on different lattice sites, $\langle p \alpha \sigma \vert q \beta \sigma' \rangle = \delta_{p q} \delta_{\alpha \beta} \delta_{\sigma \sigma'}$ (i.e.~an orthogonal TB model).

The TB model was implemented within the framework of the NEMO 3-D NanoElectronic MOdeling software, which was used to carry out the supercell calculations. \cite{Klimeck_IEEETED_2007_2}
\\ \\
\noindent
\textbf{Section S1.B: B. 14-band \textbf{k}$\cdot$\textbf{p} Hamiltonian}
\\ \\
Previous analysis has shown that it is possible and useful to derive simple continuum models that describe the perturbed band structure of GaN$_{y}$As$_{1-y}$ and GaAs$_{1-x}$Bi$_{x}$ alloys. \cite{Reilly_SST_2002,Lindsay_SSE_2003,Broderick_SST_2013} Phenomenological approaches, principally the band-anticrossing (BAC) model, have originated from interpretation of spectroscopic data and atomistic electronic structure calculations, and are widely employed as a straightforward and efficient means by which to describe the evolution of the main features of the band structure (principally the band gap and band edge energies) with alloy composition, both in bulk materials and in heterostructures.

For GaN$_{y}$As$_{1-y}$ alloys it is well established that the evolution of the CB structure can be described by a simple 2-band BAC model, in which the extended states of the GaAs host matrix CB edge undergo a composition dependent repulsive interaction (of magnitude $\beta_{\scalebox{0.6}{\textrm{N}}} \sqrt{y}$, where $\beta_{\scalebox{0.6}{\textrm{N}}}$ is the BAC coupling strength and $y$ is the N composition) with a set of localised states associated with substitutional N impurities (having energy $E_{\scalebox{0.6}{\textrm{N}}}$). \cite{Reilly_SST_2009} In the case where the N-related localised states lie energetically within the host matrix CB -- as is the case in (In)GaN$_{y}$As$_{1-y}$ \cite{Reilly_SST_2009} -- the composition dependence of the BAC interaction between these two sets of states results in a strong reduction of the alloy CB edge energy with increasing $y$.  It generally established that similar behaviour is present in GaAs$_{1-x}$Bi$_{x}$: the strong reduction (increase) and composition-dependent bowing of the band gap (VB spin-orbit-splitting energy) can be described in terms of a valence band-anticrossing (VBAC) interaction between the extended states of the GaAs VB edge, and localised impurity states associated with substitutional Bi impurities, which pushes the alloy VB edge upwards in energy with increasing $x$. Detailed analysis has demonstrated that (i) incorporation of an isolated, substitutional N (Bi) impurity in GaAs leads to the formation of $A_{1}$ ($T_{2}$) symmetric localised impurity states, which lie energetically within -- i.e.~are resonant with -- the GaAs CB (VB), and (ii) an appropriate 10-band (12-band) \textbf{k}$\cdot$\textbf{p} Hamiltonian can be derived using an extended basis set which includes these $s$-like ($p$-like) localised states, in addition to the conventional 8-band basis defined by the GaAs zone-centre Bloch states associated with the lowest energy CB, and light-hole (LH), heavy-hole (HH), and spin-split-off (SO) hole VBs. \cite{Broderick_SST_2012} Despite their simplicity, these extended \textbf{k}$\cdot$\textbf{p} models have proven to be highly effective tools for the analysis of real materials and devices. For example, theoretical models based on the 10- and 12-band models have provided significant insight into the electronic and optical \cite{Broderick_bismide_chapter_2017} properties of quantum well (QW) heterostructures, and have been used as a basis for quantitative prediction and design of key properties of electrically pumped semiconductor lasers, including the optical gain and carrier recombination rates. \cite{Broderick_IEEEJSTQE_2015}

However, formally, these (V)BAC-based models suffer from two fundamental limitations. Firstly, they consider the impact on the band structure of isolated, non-interacting impurities \textit{only}: they are hence formally applicable only to ordered alloys, thereby limiting their ability to predict the properties of real materials (which inevitably contain some degree of alloy disorder). Secondly, uncertainty surrounding the (V)BAC parameters can lead to parametric ambiguity which further reduces the predictive capability. In practice this first issue is not a significant limitation, since it is typically found that (V)BAC models provide reasonably accurate descriptions of the evolution of band properties which are not particularly sensitive to the presence of short-range alloy disorder. Instead, one must be careful in the interpretation of the results of such calculations, in the knowledge that the model will describe only band properties which are not strongly effected by N and/or Bi clustering -- e.g.~the band gap and band edge energies, \cite{Broderick_SST_2012} but \textit{not} the effective masses or Land\'{e} g factors. \cite{Broderick_PRB_2014} This second issue can, in certain circumstances, represent an impediment to the development and application of BAC models due to an inability to unambiguously determine a consistent set of band parameters (of which there can be many) from a given set of experimental or theoretical band structure data (which are typically few in number). We have circumvented this problem in general by developing a TB-based approach \cite{Usman_PRB_2011} that allows the matrix elements of the alloy Hamiltonian to be \textit{directly} calculated and parametrised in a chosen basis of crystal eigenstates and which -- via construction of the localised impurity states associated with substitutional N and/or Bi incorporation -- allows explicit evaluation of the various contributions to each matrix element -- including direct computation of the (V)BAC parameters as well as virtual crystal (conventional alloy) contributions to the band edge energies -- without the usual requirement to undertake post hoc fitting to the results of experimental measurements on GaN$_{y}$As$_{1-y}$ or GaAs$_{1-x}$Bi$_{x}$ alloys.

Turning our attention to GaN$_{y}$As$_{1-x-y}$Bi$_{x}$ alloys, the analysis summarised above suggests that a suitable basis set must contain a minimum of 14 bands: the spin-degenerate CB, LH, HH and SO bands of the GaAs host matrix (8 bands), the $A_{1}$-symmetric N-related localised states (of which there is one spin-degenerate set; 2 bands), and the $T_{2}$-symmetric Bi-related localised states (of which there are two spin-degenerate sets; 4 bands). TB calculations on ordered GaN$_{y}$As$_{1-x-y}$Bi$_{x}$ supercells indicate that the respective impact of N and Bi on the CB and VB structure are decoupled from one another. As such, the GaN$_{y}$As$_{1-x-y}$Bi$_{x}$ band structure then admits a simple interpretation in terms of perturbation of the CB and VB separately by N- and Bi-related localised states, respectively. As we demonstrate in Sec.~III B of the main paper, the 14-band model defined in this manner -- and parametrised directly from TB calculations -- provides a simple and predictive means by which to describe the band edge energies GaN$_{y}$As$_{1-x-y}$Bi$_{x}$ alloys, even in the presence of significant alloy disorder. Details of the derivation of the 14-band \textbf{k}$\cdot$\textbf{p} model can be found in Ref.~\onlinecite{Broderick_SST_2013}.

Here, we focus on the calculation of the band edge energies in pseudomorphically strained GaN$_{y}$As$_{1-x-y}$Bi$_{x}$ alloys and QWs using the 14-band \textbf{k}$\cdot$\textbf{p} Hamiltonian. At the centre of the Brillouin zone ($\textbf{k} = 0$), and in the presence of pseudomorphic strain corresponding to growth along the (001) direction, the 14-band Hamiltonian block diagonalises into decoupled sub-matrices describing the CB, HH, and LH and SO states. There are six such matrices in total, three of which are distinct as a result of spin degeneracy. The band edge energies are then given by the eigenvalues of the following spin-degenerate matrices \cite{Tomic_IEEEJSTQE_2003,Broderick_SST_2013,Broderick_SST_2015}

\begin{widetext}

\begin{eqnarray}
	H_{\scalebox{0.6}{\textrm{CB}}} &=& \left(
	\begin{array}{cc}
		E_{g}^{(0)} + \Delta E_{\scalebox{0.6}{\textrm{N}}} + \delta E_{\scalebox{0.6}{\textrm{N}}}^{\scalebox{0.7}{\textrm{hy}}} & \beta_{\scalebox{0.6}{\textrm{N}}} \sqrt{y} \\
		\beta_{\scalebox{0.6}{\textrm{N}}} \sqrt{y} & E_{g}^{(0)} + \delta E_{\scalebox{0.6}{\textrm{CB}}}^{\scalebox{0.6}{\textrm{VC}}} + \delta E_{\scalebox{0.6}{\textrm{CB}}}^{\scalebox{0.7}{\textrm{hy}}}
	\end{array}
	\right) \, ,
	\label{eq:cb_matrix} \\
	H_{\scalebox{0.6}{\textrm{HH}}} &=& \left(
	\begin{array}{cc}
		\delta E_{\scalebox{0.6}{\textrm{VB}}}^{\scalebox{0.6}{\textrm{VC}}} + \delta E_{\scalebox{0.6}{\textrm{VB}}}^{\scalebox{0.7}{\textrm{hy}}} - \delta E_{\scalebox{0.6}{\textrm{VB}}}^{\scalebox{0.7}{\textrm{ax}}} & \beta_{\scalebox{0.6}{\textrm{Bi}}} \sqrt{x} \\
		\beta_{\scalebox{0.6}{\textrm{Bi}}} \sqrt{x} & \Delta E_{\scalebox{0.6}{\textrm{Bi}}} + \delta E_{\scalebox{0.6}{\textrm{Bi}}}^{\scalebox{0.7}{\textrm{hy}}} - \delta E_{\scalebox{0.6}{\textrm{Bi}}}^{\scalebox{0.7}{\textrm{ax}}}
	\end{array}
	\right) \, ,
	\label{eq:hh_matrix} \\
	H_{\scalebox{0.6}{\textrm{LH,SO}}} &=& \left(
	\begin{array}{ccc}
		\delta E_{\scalebox{0.6}{\textrm{VB}}}^{\scalebox{0.6}{\textrm{VC}}} + \delta E_{\scalebox{0.6}{\textrm{VB}}}^{\scalebox{0.7}{\textrm{hy}}} + \delta E_{\scalebox{0.6}{\textrm{VB}}}^{\scalebox{0.7}{\textrm{ax}}} & - \sqrt{2} \, \delta E_{\scalebox{0.6}{\textrm{VB}}}^{\scalebox{0.7}{\textrm{ax}}} & \beta_{\scalebox{0.6}{\textrm{Bi}}} \sqrt{x} \\
		- \sqrt{2} \, \delta E_{\scalebox{0.6}{\textrm{VB}}}^{\scalebox{0.7}{\textrm{ax}}} & - \Delta_{\scalebox{0.6}{\textrm{SO}}}^{(0)} + \delta E_{\scalebox{0.6}{\textrm{SO}}}^{\scalebox{0.6}{\textrm{VC}}} + \delta E_{\scalebox{0.6}{\textrm{VB}}}^{\scalebox{0.7}{\textrm{hy}}} & 0 \\
		\beta_{\scalebox{0.6}{\textrm{Bi}}} \sqrt{x} & 0 & \Delta E_{\scalebox{0.6}{\textrm{Bi}}} + \delta E_{\scalebox{0.6}{\textrm{Bi}}}^{\scalebox{0.7}{\textrm{hy}}} + \delta E_{\scalebox{0.6}{\textrm{Bi}}}^{\scalebox{0.7}{\textrm{ax}}}
	\end{array}
	\right) \, ,
	\label{eq:lh_so_matrix}
\end{eqnarray}

\end{widetext}

\noindent
where $E_{g}^{(0)}$ and $\Delta_{\scalebox{0.6}{\textrm{SO}}}^{(0)}$ are the band gap and VB spin-orbit splitting energy of the GaAs host matrix, and the zero of energy has been set at the (unperturbed) GaAs VB edge. These matrices can be diagonalised to provide analytical expressions for the GaN$_{y}$As$_{1-x-y}$Bi$_{x}$ band edge energies: the lower eigenvalue of Eq.~\eqref{eq:cb_matrix} is the alloy CB edge energy, the upper eigenvalue of Eq.~\eqref{eq:hh_matrix} is the alloy HH band edge energy, and the highest and lowest eigenvalues of Eq.~\eqref{eq:lh_so_matrix} are, respectively, the alloy LH and SO band edge energies.

In Eqs.~\eqref{eq:cb_matrix} --~\eqref{eq:lh_so_matrix} $\Delta E_{\scalebox{0.6}{\textrm{N}}}$ ($\Delta E_{\scalebox{0.6}{\textrm{Bi}}}$) denotes the energy of the N- (Bi-) related localised states relative to the unperturbed GaAs CB (VB) edge: $\Delta E_{\scalebox{0.6}{\textrm{N}}} > 0$ ($< 0$) corresponds to a resonant (bound) N-related localised state lying energetically within the GaAs CB (band gap), while $\Delta E_{\scalebox{0.6}{\textrm{Bi}}} < 0$ ($> 0$) corresponds to a resonant (bound) Bi-related localised state lying energetically within the GaAs VB (band gap). The virtual crystal (conventional alloy) shifts to the CB, VB and SO band edge energies -- which, by definition, are linear in the N and Bi compositions -- are respectively defined as $\delta E_{\scalebox{0.6}{\textrm{CB}}}^{\scalebox{0.6}{\textrm{VC}}} = - \alpha_{\scalebox{0.6}{\textrm{Bi}}} \, x - \alpha_{\scalebox{0.6}{\textrm{N}}} \, y$, $\delta E_{\scalebox{0.6}{\textrm{VB}}}^{\scalebox{0.6}{\textrm{VC}}} = \kappa_{\scalebox{0.6}{\textrm{Bi}}} \, x + \kappa_{\scalebox{0.6}{\textrm{N}}} \, y$ and $\delta E_{\scalebox{0.6}{\textrm{SO}}}^{\scalebox{0.6}{\textrm{VC}}} = - \gamma_{\scalebox{0.6}{\textrm{Bi}}} \, x - \gamma_{\scalebox{0.6}{\textrm{N}}} \, y$. \cite{Broderick_SST_2013,Broderick_SST_2015}

\begin{table}[t!]
	\caption{\label{tab:14band_parameters} N- and Bi-related parameters for the 14-band Hamiltonian of GaN$_{y}$As$_{1-x-y}$Bi$_{x}$, obtained on the basis of the tight-binding supercell calculations. In GaAs$_{1-x}$Bi$_{x}$ (GaN$_{y}$As$_{1-y}$) $\Delta E$ is given relative to the unperturbed GaAs valence (conduction) band edge. Due to their $A_{1}$ symmetry, the energy of the localised states associated with isolated substitutional N impurities is not affected by axial strain: there is no associated axial deformation potential $b_{\protect\scalebox{0.6}{\textrm{N}}}$.}
	\begin{ruledtabular}
		\begin{tabular}{ccc}
			Parameter & GaAs$_{1-x}$Bi$_{x}$ & GaN$_{y}$As$_{1-y}$ \\
			\hline
			$\Delta E$ (eV) & $-0.183$ & $0.187$ \\
			$\alpha$   (eV) & $~2.63$  & $~1.55$ \\
			$\beta$    (eV) & $~1.13$  & $~2.00$ \\
			$\gamma$   (eV) & $~0.55$  & $-1.53$ \\
			$\kappa$   (eV) & $~1.01$  & $~1.36$ \\
			$a$        (eV) & $-1.11$  & $~0.83$ \\
			$b$        (eV) & $-1.71$  &  -----  \\
		\end{tabular}
	\end{ruledtabular}
\end{table}

The energy shifts due to the hydrostatic and axial components of the strain are given respectively by $\delta E_{i}^{\scalebox{0.7}{\textrm{hy}}} = a_{i} ( \epsilon_{xx} + \epsilon_{yy} + \epsilon_{zz} )$ and $\delta E_{i}^{\scalebox{0.7}{\textrm{ax}}} = - \frac{ b_{i} }{ 2 } ( \epsilon_{xx} + \epsilon_{yy} - 2 \epsilon_{zz} )$, where $i =$ CB, VB, N or Bi denotes the hydrostatic and axial deformation potentials $a_{i}$ and $b_{i}$ associated respectively with the GaAs host matrix CB and VB edges, and with the N- and Bi-related localised states. \cite{Broderick_SST_2015,Chai_SST_2015} We note that (i) for pseudomorphic strain, $\epsilon_{xx} = \epsilon_{yy}$ and $\epsilon_{zz} = - \frac{ 2 C_{12} }{ C_{11} } \epsilon_{xx}$, and (ii) by symmetry, $b_{\scalebox{0.6}{\textrm{N}}} = 0$ (since the axial component of the strain has no effect on the purely $s$-like N-related localised states). The components of the strain tensor are determined via the mismatch between the lattice constants $a(x,y)$ and $a_{\scalebox{0.6}{\textrm{S}}}$ of GaN$_{y}$As$_{1-x-y}$Bi$_{x}$ and the GaAs substrate: $\epsilon_{xx} = \frac{ a_{\scalebox{0.5}{\textrm{S}}} - a(x,y) }{ a_{\scalebox{0.5}{\textrm{S}}} }$. The lattice and elastic constants, and CB and VB edge deformation potentials, are determined for GaN$_{y}$As$_{1-x-y}$Bi$_{x}$ by interpolating linearly between those of GaN, GaAs and GaBi. \cite{Broderick_SST_2015,Vurgaftman_JAP_2003,Ferhat_PRB_2006} The N- and Bi-related parameters of the 14-band model, derived on the basis of TB supercell calculations, \cite{Broderick_SST_2013,Broderick_SST_2015} are provided in Table~\ref{tab:14band_parameters}.

For the analysis of the electronic and optical properties of QWs, the 14-band Hamiltonian is solved directly in the envelope function approximation (EFA) for each QW heterostructure using a numerically efficient reciprocal space plane wave approach. \cite{Broderick_IEEEJSTQE_2015} The QW band structure calculations are carried out in the axial approximation. In the plane wave approach the momentum matrix elements $P_{nm}^{(\widehat{e})} ( \textbf{k}_{\parallel} )$ -- between the QW eigenstates $\vert n \textbf{k}_{\parallel} \sigma \rangle$ and $\vert m \textbf{k}_{\parallel} \sigma \rangle$ at in-plane wave vector $\textbf{k}_{\parallel}$ -- are computed in reciprocal space using the general formulation due to Szmulowicz \cite{Broderick_IEEEJSTQE_2015}

\begin{equation}
	P_{nm}^{(\widehat{e})} ( \textbf{k}_{\parallel} ) = \frac{ m_{0} }{ \hbar } \sum_{\sigma, \sigma'} \langle n \textbf{k}_{\parallel} \sigma \vert \widehat{e} \cdot \nabla_{\textbf{k}} \widehat{H} \vert m \textbf{k}_{\parallel} \sigma' \rangle \, ,
	\label{eq:momentum_kdotp}
\end{equation}

\noindent
where $\widehat{e} \cdot \nabla_{\textbf{k}} \widehat{H}$ is the operator obtained by (i) taking the directional derivative of the matrix elements of the bulk \textbf{k}$\cdot$\textbf{p} Hamiltonian with respect to \textbf{k} along $\widehat{e}$, and (ii) symmetrising with respect to position-dependent material parameters and quantising $k_{z}$ in the usual manner. Using Eqs.~\eqref{eq:momentum_tb} and~\eqref{eq:momentum_kdotp} we compute the optical transition strength (in units of energy) directly for each structure in terms of the zone-centre momentum matrix element as $\frac{ 2 m_{0} }{ \hbar^{2} } \vert P_{nm}^{ ( \widehat{e} ) } (0) \vert^{2}$, where $\widehat{e} = \widehat{x}$ and $\widehat{z}$ for transverse electric- (TE-) and transverse magnetic- (TM-) polarised transitions, respectively. We note that this approach to calculating the optical transition strengths directly employs the supercell Hamiltonian and computed eigenstates for a given structure, meaning that the full effects of N- and Bi-induced hybridisation are explicitly accounted for in the analysis of the optical properties. \cite{Usman_APL_2014, Broderick_IEEEJSTQE_2015}
\\ \\


\noindent
\textbf{Section S2: Dilute doping limit: impact of co-alloying N and Bi on the GaAs electronic structure}
\\ \\
\noindent
The interactions of individual substitutional N and Bi atoms with the host matrix states in the GaN$_{y}$As$_{1-y}$ and GaAs$_{1-x}$Bi$_{x}$ alloys, respectively, have been well studied in the literature. \cite{Reilly_SST_2009,Usman_PRB_2011} An isolated N atom introduces a resonant impurity state above the CB edge in GaAs \cite{Reilly_SST_2009}, while an isolated Bi atom introduces a impurity state lying below the GaAs VB edge in energy. \cite{Usman_PRB_2011} However, no quantitative analysis has been undertaken to date to quantify the extent of any interaction between the localised impurity states associated with N and Bi when both are incorporated substitutionally into GaAs. Since their individual behaviours are characterized by BAC interactions in the CB and VB, it is pertinent to investigate the degree to which this behaviour remains intact in the quaternary GaN$_{y}$As$_{1-x-y}$Bi$_{x}$ alloy. Here, we systematically analyse the interaction between these localised states by placing single substitutionaly N and Bi atoms inside a large 4096-atom Ga$_{2048}$N$_{1}$As$_{2046}$Bi$_{1}$ supercell, and probe the interaction as a function of the spatial separation of these impurities. We begin by placing the N and Bi atoms sufficiently far apart that the interaction between their associated localised impurity states is minimal, and then gradually bring them closer together, finally studying the case where the N and Bi atoms are located at adjacent sites on the anion sublattice (i.e.~so that they are second-nearest neighbours, sharing a common Ga nearest neighbour). This latter case is that of maximum spatial overlap between the N- and Bi-related localised impurity states, and hence represents the case in which the interaction between these states would be maximised in a substitutional GaN$_{y}$As$_{1-x-y}$Bi$_{x}$ alloy. As described in the main paper, we analyse the character of this interaction by computing the fractional GaAs $\Gamma$ character of the alloy CB and VB edge states in each supercell.

Figure~\ref{fig:anion_environments} contains schematic illustrations of the different local neighbour environments considered in this analysis. The relative positions of the N and Bi atoms in each case are defined in Sec.~III A of the main text. The distinct local neighbour environments considered are (a) an unperturbed GaAs supercell, as well as a GaAs supercell containing (b) a single N impurity, (c) a single Bi impurity, and GaAs supercells in which a single Bi impurity is placed with respect to the N atom at the (d) third-closest, (e) second-closest, and (f) closest sites on the anion sub-lattice. This final case is that referred to above, whereby the N and Bi atoms are second-nearest neighbours sharing a common Ga nearest neighbour. The labelling scheme used to denote these distinct local neighbour environments is defined in Sec.~III A of the main paper.

Figure~\ref{fig:gamma_character_ordered} shows the fractional GaAs CB and VB edge $\Gamma$ character $G_{\Gamma} (E)$ spectra for the ternary Ga$_{2048}$N$_{1}$As$_{2047}$ and Ga$_{2048}$As$_{2047}$Bi$_{1}$ supercells. These spectra clearly indicate hybridisation between the GaAs CB (VB) edge states and a resonant state associated with the substitutional N (Bi) atoms, lying energetically within the CB (VB) of the GaAs host matrix semiconductor. This hybridisation is the signature of the (V)BAC interactions described in Sec.~S1 above, and is consistent with the established trends described in the references provided therein. For the Ga$_{2048}$N$_{1}$As$_{2047}$ in Fig.~S2 (a) we calculate the the alloy CB edge state has 87.4\% GaAs CB edge character and is shifted downwards in energy by 19 meV with respect to the unperturbed GaAs CB edge. Almost the entirety of the remainder of the GaAs CB edge character is calculated to reside on a state lying at an energy of approximately 1.64 eV, and which is highly localised about the N atomic site. We compute that the N localised resonant state if of energy 1.62 eV in this supercell -- i.e.~lying approximately 190 meV above the room temperature GaAs CB edge. Figure S2 (b) demonstrates that the VB edge states in Ga$_{2048}$N$_{1}$As$_{2047}$ are virtually unchanged compared to those in a pure Ga$_{2048}$As$_{2048}$ supercell, with the alloy VB edge states having $> 99.9$\% overlap with the VB edge states of the unperturbed GaAs host matrix, reflecting that while N incorporation significantly perturbs the CB structure, its impact on the VB structure is minimal.

In Fig.~S2(d) the Ga$_{2048}$As$_{2047}$Bi$_{1}$ supercell we observe complementary behaviour: we find that the majority ($> 96$\%) of the GaAs VB edge character resides on the alloy VB edge states, which are shifted upwards in energy compared to those in the GaAs host matrix. The remainder of the GaAs VB edge character primarily resides on a single alloy VB state, lying $\sim 100$ meV below the unperturbed GaAs VB edge in energy, with further smaller amounts projected onto lower lying VB states (off scale in Fig.~S2 (d)). Again, this hybridisation describes the presence of BAC behaviour, this time resulting from coupling between the VB edge states of the GaAs host matrix, and impurity states that are resonant with the GaAs VB and strongly localised about the Bi atomic site. Fig.~S3 (c) highlights that the calculated overlap between the alloy CB edge state and the CB edge state of unperturbed GaAs is $> 99.9$\%, reflecting that Bi incorporation primarily impact the VB structure while having minimal impact on the CB structure.

Having understood the key trends in the ternary supercell calculations, we now turn our attention to the corresponding calculations for the quaternary Ga$_{2048}$N$_{1}$As$_{2046}$Bi$_{1}$ supercells containing both N and Bi, and analyse the changes in the calculated fractional GaAs CB and VB edge $\Gamma$ character brought about by co-alloying N and Bi. Table I in the main text provides the band edge energies calculated for each of these supercells. Examining the fractional GaAs CB edge $\Gamma$ character Figs.~S3 (a), (c) and (e), we note that the trends observed in the presence of Bi remain relatively unchanged from those calculated for the ternary Ga$_{2048}$N$_{1}$As$_{2047}$ supercell. This indicates that the impact of N on the CB structure in GaN$_{y}$As$_{1-x-y}$Bi$_{x}$ remains close to that in the ternary GaN$_{y}$As$_{1-y}$ alloy when Bi is incorporated, providing further confirmation of the trends identified in the main paper: Bi incorporation produces only minor quantitative changes to the nature of the CB structure, with the overall character remaining qualitatively unchanged, even in the case when substitutional N and Bi atoms share common Ga nearest neighbours. Turning our attention to the VB structure, we again note the same general trend: the overall character of the alloy VB edge states is qualitatively similar to that in the ternary GaAs$_{1-x}$Bi$_{x}$ alloy, with relatively minor quantitative changed brought about by co-alloying with N. The primary feature revealed by the calculated fractional GaAs VB edge $\Gamma$ character in Figs.~S3 (b), (d) and (f) is that the presence of N tends to lift degeneracy of the LH- and HH-like VB edge states. This effect is associated with the fact that the different local neighbour environments containing both N and Bi represent disorder alloy microstructure in which the translational symmetry of the crystal lattice is broken, leading to a breakdown of the $T_{d}$ point group symmetry of the GaAs lattice. This lack of isotropy in the local crystal structure, which is compounded by the large local relaxation of the crystal lattice about complexes of spatially proximate N and Bi atoms, is then manifested in the lifting of the degeneracy of the $p$-like alloy VB edge states due to the associated non-equivalence of the [001], [010] and [001] crystal directions. This lifting of the VB edge energy is qualitatively identical to that we have calculated previously in disordered GaAs$_{1-x}$Bi$_{x}$ alloys, but in quantitative terms tends to be larger in magnitude in GaN$_{y}$As$_{1-x-y}$Bi$_{x}$ due to the aforementioned large local relaxation of the crystal lattice in cases where N and Bi atoms are located close to one another.
\\ \\


\noindent
\textbf{Section S3: Electron and hole probability density plots for GaBiAs/GaAs QWs:}
\\ \\
\noindent
Here we consider the electronic structure of the N-free GaAs$_{1-x}$Bi$_{x}$/GaAs QWs having $x = 6.25$ and 9\% (structures 1 and 2 of Secs.~III C and III D in the main paper). Figures~\ref{fig:envelope_functions_GaAsBi} (a) and (b) respectively show the probability density along the (001) direction, associated with the lowest energy electron state $e1$ (upper row) and highest energy hole state $h1$ (lower row) in structures 1 and 2, for five representative supercells containing distinct random distributions (RDs) of substitutional N and Bi atoms. Solid and dashed red lines respectively denote the probability density projected to cations and anions, calculated using the TB method at a fixed position $z$ along (001) by summing over the probability density associated with all atoms in the plane perpendicular to (001). The plots from \textbf{k}$\cdot$\textbf{p} model are shown in the Fig. 3 of the main text. Comparison of the TB and \textbf{k}$\cdot$\textbf{p} calculations suggest that the electron states in GaAs$_{1-x}$Bi$_{x}$/GaAs QWs are broadly similar in nature to those in conventional QWs -- e.g.~In$_{x}$Ga$_{1-x}$As/GaAs -- since the electron eigenstates are (i) well described in terms of envelope functions which vary slowly and smoothly with position, (ii) effectively insensitive to the presence of underlying short-range alloy disorder, and (iii) relatively insensitive to changes in Bi composition (in this case, from $x = 6.25$ to 9\% between structures 1 and 2). This is consistent with the expected behaviour for GaAs$_{1-x}$Bi$_{x}$ alloys and heterostructures: Bi incorporation strongly perturbs the VB while leaving the CB relatively unaffected, with the evolution of the latter readily captured by conventional alloy descriptions. We note also that in all supercells considered $e1$ is localised within the GaAs$_{1-x}$Bi$_{x}$ layer, reflecting the appreciable type-I GaAs$_{1-x}$Bi$_{x}$/GaAs CB offset $\Delta E_{\scalebox{0.6}{\textrm{CB}}}$. \cite{Usman_PRB_2011,Broderick_SST_2013,Broderick_SST_2015,Usman_APL_2014} This is confirmed by the calculated increase in the $e1$ probability density in the centre of the QW in structure 2, reflecting the increase in $\Delta E_{\scalebox{0.6}{\textrm{CB}}}$ ($\sim - \alpha_{\scalebox{0.6}{\textrm{Bi}}} \, x + \delta E_{\scalebox{0.6}{\textrm{CB}}}^{\scalebox{0.6}{\textrm{hy}}}$) brought about by increasing the Bi composition from that in structure 1.

By contrast, the $h1$ probability density (lower row, Figs.~\ref{fig:envelope_functions_GaAsBi} (a) and (b)) calculated for structures 1 and 2 using the TB method is strongly perturbed compared to that calculated using the \textbf{k}$\cdot$\textbf{p} method in the EFA. While the $h1$ probability density is, as expected, confined within the Bi-containing layer and primarily localised at anion sites, there is a significant departure from the smooth, envelope function-like spatial variation observed for the $e1$ eigenstates. In contrast to the probability density associated with $e1$, which is spread across the extent of the QW, for $h1$ we observe extremely strong localisation of the probability density at various different locations within the QW, with strong dependence of this localisation on the precise short-range alloy disorder present in the QW. We have previously identified that this unusual behaviour has its origins in Bi-related alloy disorder, specifically in the formation of pairs and larger clusters of Bi atoms in a realistic GaAs$_{1-x}$Bi$_{x}$ alloy. \cite{Usman_PRB_2013} As described above, these pairs and clusters create a full distribution of localised states lying close in energy to the VB edge, which then hybridise strongly with extended GaAs VB states leading to a multiplicity of VB edge states having significant localised character (low Bloch character). Further analysis reveals that this strong spatial localisation of the hole eigenstates typically occurs about the largest Bi clusters present in a given alloy (reflected here in the sense that a large TB calculated probability density reflects the presence of Bi pairs and/or clusters in the plane at that value of $z$). The precise distribution of Bi-related localised states depends closely on the precise short-range alloy disorder present in the GaAs$_{1-x}$Bi$_{x}$ layers in these QW structures, thereby accounting for the observed strong variation of the localisation of the $h1$ eigenstates at fixed compositions for structures containing different RDs of Bi atoms. Comparing the $h1$ probability density from the TB calculations with that obtained using the 14-band \textbf{k}$\cdot$\textbf{p} model in the EFA (shown in Fig. 4 of the main paper), we conclude that (i) the strongly perturbative impact of Bi incorporation and alloy disorder on the VB structure in GaAs$_{1-x}$Bi$_{x}$ leads to a strong breakdown of the envelope function description of the hole eigenstates in GaAs$_{1-x}$Bi$_{x}$/GaAs heterostructures, and (ii) this breakdown of the EFA for hole eigenstates is a generic feature of the electronic properties GaAs$_{1-x}$Bi$_{x}$/GaAs in the sense that it is intrinsic and effectively insensitive to the Bi composition.
\\ \\


\noindent
\textbf{Section S5: Probability densities in GaN$_{y}$As$_{1-x-y}$Bi$_{x}$/GaAs quantum wells}
\\ \\
\noindent
In the main paper we present probability densities calculated for the same 8 nm thick GaN$_{0.025}$As$_{0.9125}$Bi$_{0.0625}$/GaAs and GaN$_{0.01}$As$_{0.90}$Bi$_{0.09}$/GaAs QWs (structures 3 and 4, respectively) using the TB method, for selected supercells containing distinct RDs of N and Bi atoms substituted on the anion sublattice in the well region (cf.~Fig.~\ref{fig:qw_structure}). Figures~\ref{fig:envelope_functions_GaNAsBi} (a) and (b) respectively present probability densities calculated for these two QWs, for several additional RDs in each case. The line types are as in Fig.~2 of the main paper. The calculated trends are qualitatively similar here to those discussed for the other RDs in the main paper and, as such, we refer the reader to Sec.~III C of the main paper for a discussion of the associated trends.


  \begin{figure*}[t!]
  	\includegraphics[width=0.8\columnwidth]{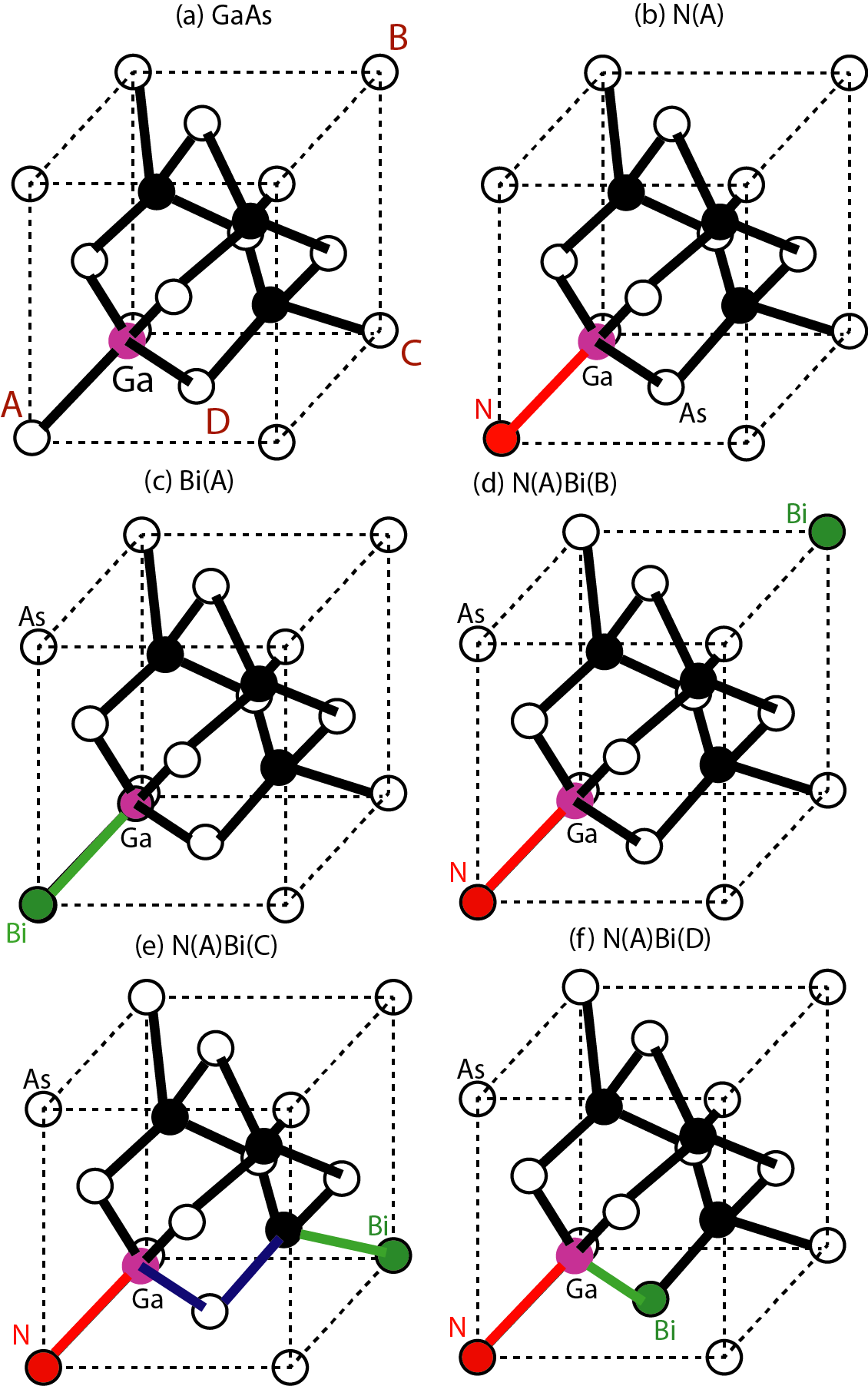}
  	\caption{Schematic illustration of a single, simple cubic 8-atom GaAs unit cell within the 4096 atom supercells considered. N and Bi atoms are substituted at specific As sites A, B, C and D on the anion sublattice (Ga and As atoms depicted respectively in black and white). (a) A pure Ga$_{2048}$As$_{2048}$ supercell, with the location of the anion lattice sites A -- D indicated. (b) A Ga$_{2048}$N$_{1}$As$_{2047}$ supercell containing an isolated substitutional N impurity (depicted in red) at the anion lattice site A (denoted by GaAs:N$_{\protect\scalebox{0.6}{\textrm{A}}}$). (c) A Ga$_{2048}$As$_{2047}$Bi$_{1}$ supercell containing an isolated substitutional Bi impurity (depicted in green) at the anion lattice site A (denoted by GaAs:Bi$_{\protect\scalebox{0.6}{\textrm{A}}}$). (d) A Ga$_{2048}$N$_{1}$As$_{2046}$Bi$_{1}$ supercell containing single substitutional N and Bi impurities at the respective anion lattice sites A and B (denoted by GaAs:N$_{\protect\scalebox{0.6}{\textrm{A}}}$Bi$_{\protect\scalebox{0.6}{\textrm{B}}}$), so that the N and Bi atoms are third-nearest neighbours. (e) A Ga$_{2048}$N$_{1}$As$_{2046}$Bi$_{1}$ supercell containing single substitutional N and Bi impurities at the respective anion lattice sites A and C (denoted by GaAs:N$_{\protect\scalebox{0.6}{\textrm{A}}}$Bi$_{\protect\scalebox{0.6}{\textrm{C}}}$), so that the N and Bi atoms are second-nearest neighbours. (f) A Ga$_{2048}$N$_{1}$As$_{2046}$Bi$_{1}$ supercell containing single substitutional N and Bi impurities at the respective anion lattice sites A and D (denoted by GaAs:N$_{\protect\scalebox{0.6}{\textrm{A}}}$Bi$_{\protect\scalebox{0.6}{\textrm{D}}}$), so that the N and Bi atoms are second-nearest neighbours.}
  	\label{fig:anion_environments}
  \end{figure*}

  \begin{figure*}[t!]
  	\includegraphics[width=1.5\columnwidth]{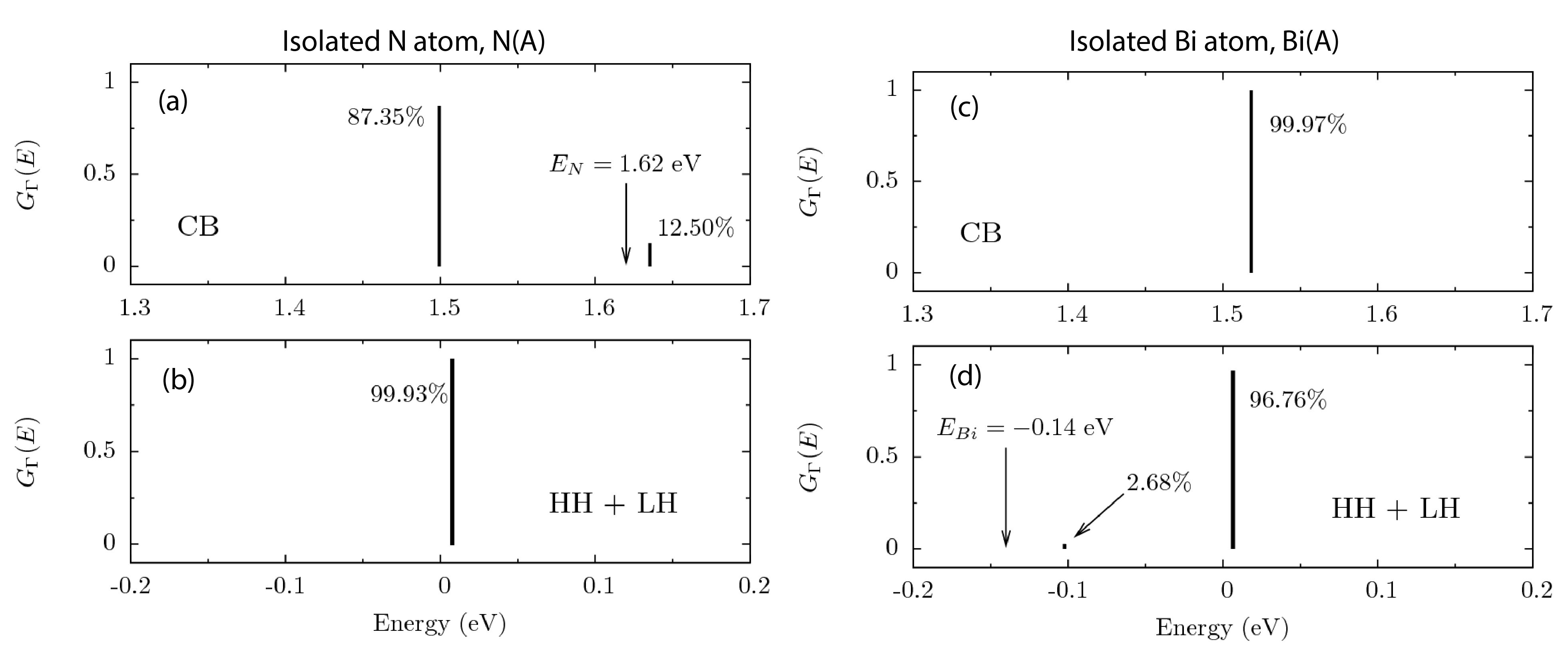}
  	\caption{Calculated GaAs fractional $\Gamma$ character spectra $G_{\Gamma} (E)$ for the CB and VB edge states in ordered 4096-atom Ga$_{2048}$N$_{1}$As$_{2047}$ and Ga$_{2048}$As$_{2047}$Bi$_{1}$ supercells. (a) and (b) show $G_{\Gamma} (E)$ associated respectively with the GaAs CB and VB (i.e.~combined LH and HH) edge states in Ga$_{2048}$N$_{1}$As$_{2047}$. (c) and (d) show $G_{\Gamma} (E)$ associated respectively with the GaAs CB and VB edge states in Ga$_{2048}$As$_{2047}$Bi$_{1}$. Band-anticrossing interactions associated with N- (Bi-) related localised resonant states are clearly visible in $G_{\Gamma} (E)$ calculated for the Ga$_{2048}$N$_{1}$As$_{2047}$ CB (Ga$_{2048}$As$_{2047}$Bi$_{1}$ VB), consistent with our previous calculations in Refs.~\cite{Lindsay_SSC_2001} and~\cite{Usman_PRB_2011}.}
  	\label{fig:gamma_character_ordered}
  \end{figure*}

 \begin{figure*}[t!]
  	\includegraphics[width=2\columnwidth]{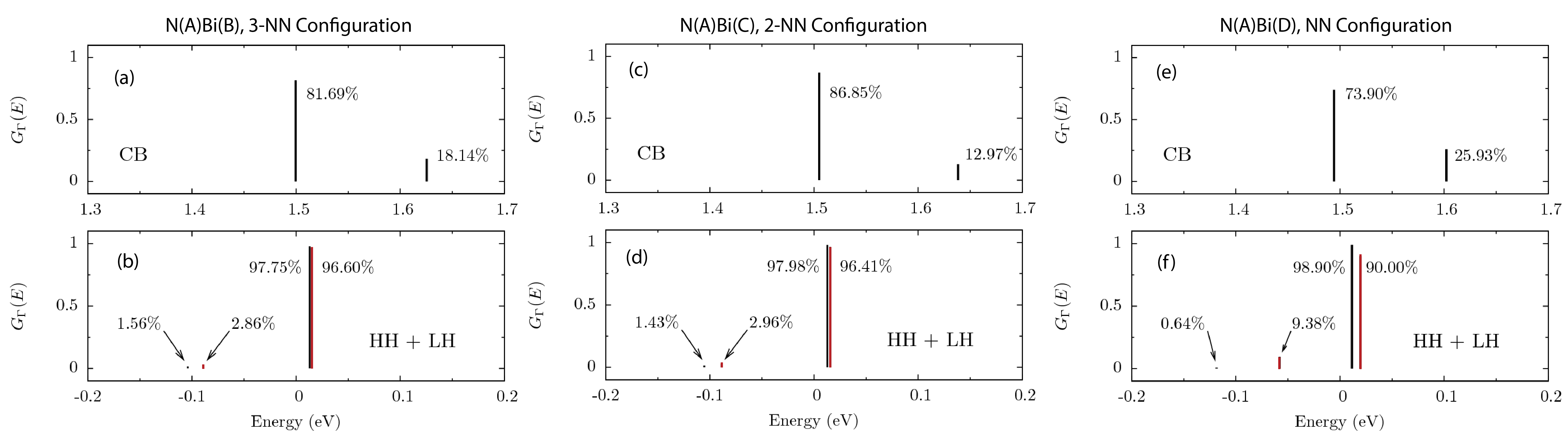}
  	\caption{Calculated GaAs fractional $\Gamma$ character spectra $G_{\Gamma} (E)$ for the CB and VB edge states in 4096-atom Ga$_{2048}$N$_{1}$As$_{2046}$Bi$_{1}$ supercells containing one N and one Bi impurity arranged in different local configurations (cf.~Fig.~\ref{fig:anion_environments}). (a) and (b) show $G_{\Gamma} (E)$ associated respectively with the GaAs CB and VB (i.e.~combined LH and HH) edge states in the GaAs:N$_{\protect\scalebox{0.6}{\textrm{A}}}$Bi$_{\protect\scalebox{0.6}{\textrm{B}}}$ supercell. (a) and (b), (c) and (d), and (e) and (f) show, respectively, $G_{\Gamma} (E)$ associated respectively with the GaAs CB and VB (i.e.~combined HH and LH) edge states in the GaAs:N$_{\protect\scalebox{0.6}{\textrm{A}}}$Bi$_{\protect\scalebox{0.6}{\textrm{B}}}$, GaAs:N$_{\protect\scalebox{0.6}{\textrm{A}}}$Bi$_{\protect\scalebox{0.6}{\textrm{C}}}$, and GaAs:N$_{\protect\scalebox{0.6}{\textrm{A}}}$Bi$_{\protect\scalebox{0.6}{\textrm{D}}}$ supercells. In moving the Bi atom from the I to E to B anion lattice site (cf.~Fig.~\ref{fig:anion_environments} the Bi atom is brought closer to the N atom at anion site A. The reduced symmetry in these supercells lifts the degeneracy of the LH- and HH-like Bi-related and VB edge states, with the energy splitting between these states increasing as the Bi atom is brought closer to the N site, due to larger local relaxation of the crystal lattice. }
  	\label{fig:gamma_character_disordered}
  \end{figure*}

 \begin{figure*}[t!]
  	\includegraphics[width=0.6\columnwidth]{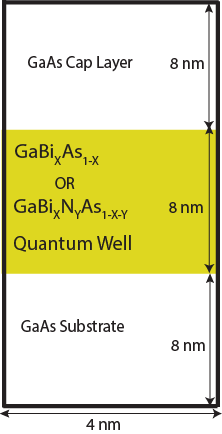}
  	\caption{Schematic illustration of the structures considered in the analysis of the electronic and optical properties of GaN$_{y}$As$_{1-x-y}$Bi$_{x}$/GaAs quantum wells. We consider an 8 nm thick quantum well surrounded by 8 nm thick GaAs barriers, giving a total length of 24 nm along the (001) growth direction and a separation of 16 nm between image quantum wells associated with the Born von Karman boundary conditions. The lateral dimensions are taken to be 4 nm along the (100) and (010) in-plane directions. In the tight-binding calculations this corresponds to a supercell of volume 384 nm$^{3}$ containing a total of 24,576 atoms. The same geometry along (001) is employed in the 14-band \textbf{k}$\cdot$\textbf{p} calculations with the exception that the calculation proceeds in one dimension only, so that the lateral dimensions are inconsequential.}
  	\label{fig:qw_structure}
  \end{figure*}

 \begin{figure*}[t!]
  	\includegraphics[width=1.00\textwidth]{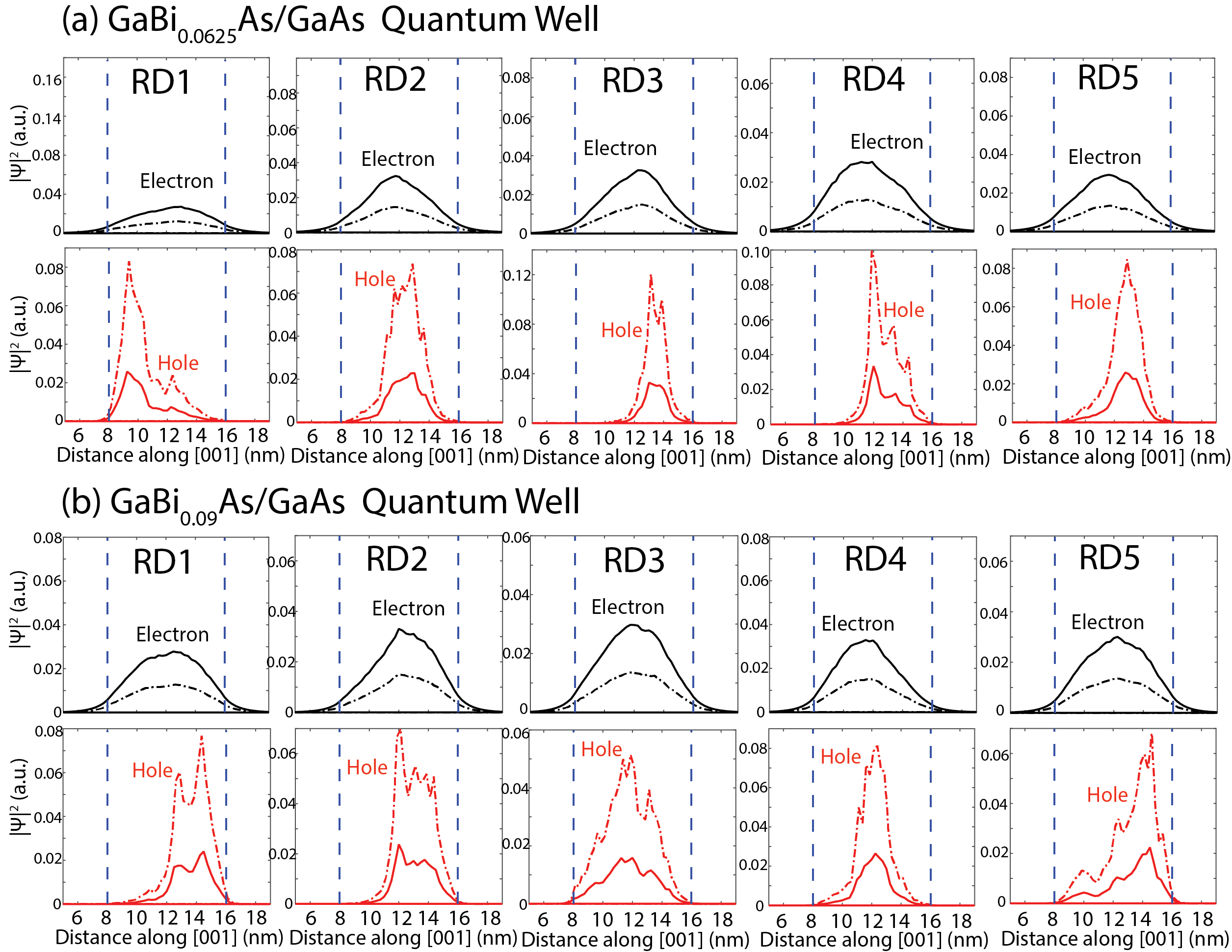}
  	\caption{Probability density associated with the lowest energy conduction electron state (upper row) and highest energy valence hole state (lower row) in N-free (a) GaAs$_{0.9375}$Bi$_{0.0625}$/GaAs ($x = 6.25$\%), and (b) GaAs$_{0.91}$Bi$_{0.09}$/GaAs ($x = 9$\%) quantum wells (cf.~Fig.~\ref{fig:qw_structure}). Solid (dash-dotted) black lines and solid (dash-dotted) red lines show, respectively, the electron and hole probability density at cation (anion) sites calculated using the $sp^{3}s^{*}$ tight-binding model, obtained at each position $z$ along the growth direction by summing over the probability densities associated with each atom in the plane. Vertical dashed black lines denote the well/barrier interfaces. The tight-binding calculations were performed for five different random distributions (RDs), corresponding to structures with fixed $x$ and $y$ in which the N and Bi atoms were substituted at randomly selected sites on the anion sublattice. It can be seen clearly that Bi-related alloy disorder strongly perturbs the VB leading to strong hole localisation, but that the CB is relatively unaffected with the electron probability densities retaining the smooth envelope function-like behaviour.}
  	\label{fig:envelope_functions_GaAsBi}
  \end{figure*}

  \begin{figure*}[t!]
  	\includegraphics[width=1.00\textwidth]{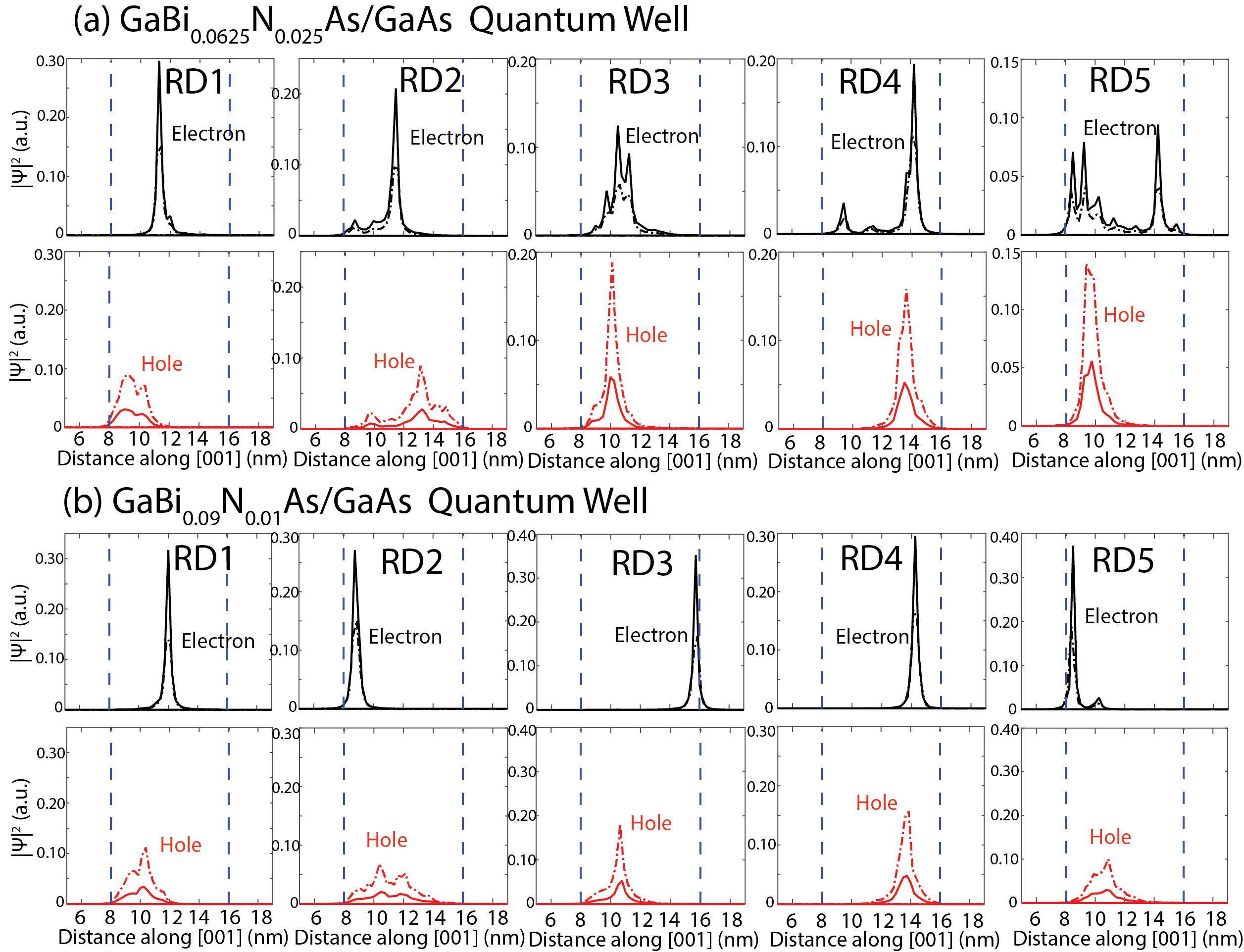}
  	\caption{Probability density associated with the lowest energy CB electron state ($e1$; upper row) and highest energy VB hole state ($h1$; lower row) in 8 nm thick (a) GaN$_{0.025}$As$_{0.9125}$Bi$_{0.0625}$/GaAs ($x = 6.25$\%, $y = 2.5$\%), and (b) GaN$_{0.01}$As$_{0.90}$Bi$_{0.09}$/GaAs ($x = 9$\%, $y = 1$\%) QWs. Solid (dash-dotted) black lines and solid (dash-dotted) red lines respectively denote the $e1$ and $h1$ probability density at cation (anion) sites calculated using the TB model, obtained at each position $z$ along the [001] growth direction by summing over the probability densities associated with each atom in the plane. Dashed blue lines denote the well/barrier interfaces. The TB calculations were performed for five supercells containing different random spatial distributions (RDs) of substitutional N and Bi atoms on the anion sublattice.}
  	\label{fig:envelope_functions_GaNAsBi}
  \end{figure*}

\newpage
\clearpage

\bibliographystyle{apsrev}  

\end{document}